\documentclass[twocolumn,aps,showpacs,superscriptaddress,tightenlines,a4paper]{revtex4}

\usepackage{amssymb}
\usepackage{amsmath}
\usepackage{bbold}
\usepackage{mathrsfs}
\usepackage{paralist}
\usepackage{graphicx}
\usepackage{xcolor}

\begin{document}
%
%
    \title{Radiation pressure, moving media, and multilayer systems.}
%
%
    \author{S. A. R. Horsley}
    \affiliation{School of Physics and Astronomy, University of St Andrews,
North Haugh, St Andrews, KY16 9SS, UK}
    \email{sarh@st-andrews.ac.uk}
    \affiliation{European Laboratory for Nonlinear Spectroscopy, Sesto
Fiorentino, Italy}
    \author{M. Artoni}
    \affiliation{European Laboratory for Nonlinear Spectroscopy, Sesto
Fiorentino, Italy}
    \affiliation{Department of Physics and Chemistry of Materials \& CNR-IDASC
Sensor Lab,
Brescia University, Brescia, Italy}
    \author{G. C. La Rocca}
    \affiliation{Scuola Normale Superiore and CNISM, Pisa, Italy}
%
%
    \begin{abstract}
 A general theory of optical forces on moving bodies is here developed in terms of generalized/\(4\times4\) transfer and scattering matrices.
        Results are presented for a planar dielectric multilayer of arbitrary refractive index placed in an otherwise empty space and moving \textit{parallel} and \textit{perpendicular} to the slab-vacuum interface. In both regimes of motion the resulting force comprises lateral and normal velocity--dependent components which may depend in a subtle way on the Doppler effect and TE-TM polarization mixing. For lateral displacements in particular,  polarization mixing, which is here interpreted as an effective magneto--electric effect due to the reduced symmetry induced by the motion of the slab, gives rise to a velocity dependent force contribution that is sensitive to the phase difference between the two polarizations amplitudes. This term gives rise to a rather peculiar optical response on the moving body and specific cases are illustrated for incident radiation of arbitrarily directed linear polarization. 
        The additional force due to polarization mixing may cancel to first order in \(V/c\) with the first order Doppler contribution yielding an overall vanishing of the velocity-dependent  component of the force on the body.  The above findings bare some relevance to modern developments of nano-optomechanics as well as to the problem of a frictional component to the Casimir force.       
    \end{abstract}
%
%
    \maketitle

\section{Introduction}
    \par
    Although Minkowski derived the basic equations for describing the electrodynamics of moving media in 1908~\cite{minkowski1908}, it is clear that there is still a good deal to be done to understand the effect of the electromagnetic field on the motion of a dielectric body.  While the Abraham--Minkowskii debate~\cite{leonhardt2006,pfeifer2007,barnett2010} is the most persistent instance of this, a more tractable problem can be found within the recent controversy surrounding the existence of a frictional component to the Casimir force~\cite{pendry1997,volokitin1999,philbin2009,pendry2010a,leonhardt2010a,pendry2010b,volokitin2010}.  Although a quantum phenomenon, much of the initial debate was concerned with classical radiation pressure.  Namely, what is the electromagnetic force on a dielectric medium in motion?  The concern was with the relative importance of the relativistic transformation of polarization, versus that of frequency (the Doppler shift) to the force exerted by the electromagnetic field on a moving medium~\cite{pendry2010a,leonhardt2010a,pendry2010b}.
    \par
    A further motivation for the current investigation comes from the remarkable recent progress in optomechanics (see~\cite{florian-physics}) which uses the motional effects of radiation pressure to cool mechanical degrees of freedom to sub--Kelvin temperatures~\cite{metzger2004,corbitt2007,schliesser2008}, and even close to the quantum ground state~\cite{riviere2011}.  Relevant to this is work on `radiation damping'~\cite{braginski1967,matsko1996} in highly dispersive systems such as photonic crystals~\cite{karrai2008}, and atomic multilayers~\cite{horsley2011}.  This paper adds to other recent work on velocity dependent forces in optomechanics~\cite{xuereb2009}, generalizing such treatments to an arbitrary direction of motion.  It is also worth mentioning that our findings may be applicable within the field of metamaterials (see for example~\cite{tretyakov2009,capolino2009,leonhardt2010b}), where some aspects of motion can be mimicked with a suitably engineered magnetoelectric response (however, see~\cite{horsley2011b}), and where optical properties such as extreme sensitivity to polarization (see section~\ref{Vy-section}) may be achieved with structured surfaces~\cite{lousse2004}. 
    \par
    Here we focus on the theory of radiation pressure for planar media in relative motion, and show how the motion leads to velocity dependent forces that always have two distinct origins; one contribution coming from the mixing of s and p polarizations by the motion; and the other more familiar effect coming from the Doppler shifted frequency that appears within the rest frame reflection and transmission coefficients.  We find that whereas the latter effect may be amplified through an increase in the dispersion of the medium, the former may be amplified through an increase in the sensitivity of the medium to the polarization at small angles of incidence.  We quantify the relative importance of these two effects and show situations where either contribution can be significant.
    \par
    In section \ref{tmsec} we outline the general theory of radiation pressure for planar media, where the response is characterized in terms of scattering, or transfer matrices.  Section \ref{mmsec} then applies this to the case of moving media, illustrating that the motion can be understood in terms of a linear transformation of the scattering matrix.  We comment briefly on the likely implications of our findings to the vacuum force experienced by two dielectric plates in lateral motion.
    	\par
    \section{Calculating radiation pressure using the transfer matrix\label{tmsec}}
    \par
     Optical forces on a body are assessed by calculating either the net Lorentz force experienced by the body, 
  
    \begin{equation}
    \frac{d\boldsymbol{P}}{dt}=\int[\rho\boldsymbol{E}+\boldsymbol{j}\times\boldsymbol{B}]d^{3}\boldsymbol{x}, 
    \end{equation}
    where \(\rho=-\boldsymbol{\nabla}\boldsymbol{\cdot}\boldsymbol{P}\) and \(\boldsymbol{j}=\boldsymbol{\nabla}\boldsymbol{\times}\boldsymbol{M}+\partial\boldsymbol{P}/\partial t\) (c.f.~\cite{mkrtchian2003,loudon2006}), or the rate of momentum being lost from the field in the space outside the body.  The rate of momentum leaving the electromagnetic field can be written as an integral over the surface of  the energy--momentum tensor~\cite{volume2,volume8,genet2003},
      \begin{align}
        T^{\mu\nu}&=\left(
                \begin{matrix}
\mathscr{E}_{\text{\tiny{F}}}&c\boldsymbol{\boldsymbol{\mathcal{P}}}_{\text{\tiny{F}}}\\
c\boldsymbol{\mathcal{P}}_{\text{\tiny{F}}}&-\boldsymbol{\sigma}_{\text{\tiny{F}
}}
                \end{matrix}
            \right)\nonumber\\
            &=\left(
                \begin{matrix}
\frac{\epsilon_{0}}{2}\left(\boldsymbol{E}^{2}+c^{2}\boldsymbol{B}^{2}\right)&c\epsilon_{0}\boldsymbol{E}
\times\boldsymbol{B}\\
c\epsilon_{0}\boldsymbol{E}\times\boldsymbol{B}&\mathbb{1}_{3}\mathscr{E}_{\text{\tiny{F}}}-\epsilon_{0}[\boldsymbol{E}\otimes\boldsymbol{E
}+c^{2}\boldsymbol{B}\otimes\boldsymbol{B}]
                \end{matrix}
            \right)\nonumber\\
            \label{energy-momentum-tensor}
    \end{align}
 where $\mathscr{E}_{\text{\tiny{F}}}$ and $\boldsymbol{{\mathcal{P}}}_{\text{\tiny{F}}}$ are the energy and momentum density, respectively, and $\boldsymbol{\sigma}_F$ is the Maxwell stress tensor.  In this paper we choose to work in terms of the energy--momentum tensor, for it allows the optical force to be straightforwardly written in terms of the medium reflection and transmission coefficients.   Note that throughout it is assumed that the region outside of the material has \(\epsilon\sim\epsilon_{0}\), and \(\mu\sim\mu_{0}\), so that we may avoid possible ambiguities in the form of the energy momentum tensor within the surrounding space~\cite{leonhardt2006,barnett2010}.
%
%
\subsection{The energy-momentum tensor for a plane wave reflecting from a surface.}  
   \par
\begin{figure}[hc]
 \begin{centering}
	\includegraphics[width=8.3cm]{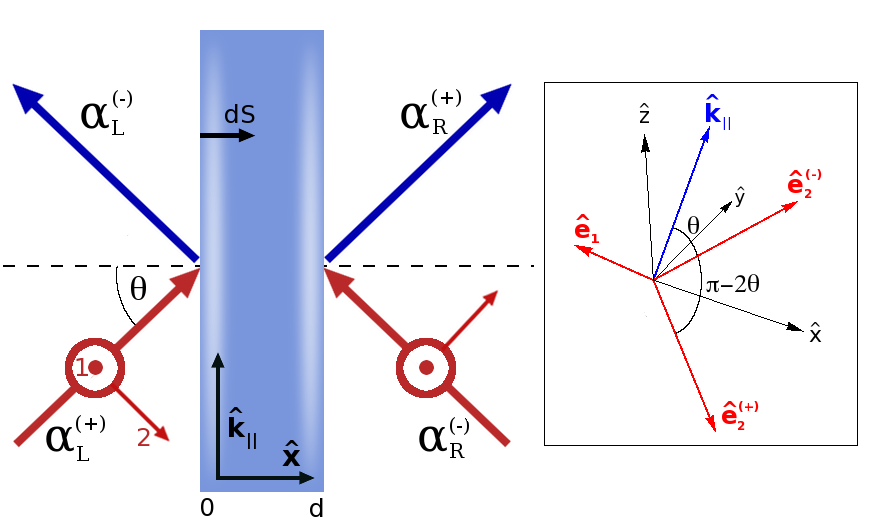}
\end{centering}
	\caption{
	Geometry of the slab interfaces showing the input (red) field amplitudes $\alpha_{\text{\tiny{L}}}^{(+)}$ and $\alpha_{\text{\tiny{R}}}^{(-)}$ and the output (blue) amplitudes $\alpha_{\text{\tiny{L}}}^{(-)}$ and $\alpha_{\text{\tiny{R}}}^{(+)}$.  
	The slab surface normal is along $\pm\hat{\boldsymbol{x}}$ while the orientation of the two polarizations (\(1,2\)) are indicated in the input amplitudes.  The inset shows the unit polarization vectors (\ref{pol1}--\ref{pol2}) and their relative orientation in terms of the angle defined in (\ref{angles}).\label{figure-1}}
    \end{figure}

    For a field with a harmonic time dependence \(\boldsymbol{E}_{\omega}(\mathbf{x})e^{-i\omega t}$ and $\boldsymbol{B}_{\omega}(\mathbf{x})e^{-i\omega t}\), observed over a time period \(\Delta t\), where \(1/\omega\ll\Delta t\), we can deal with the time average of (\ref{energy-momentum-tensor}) over many optical cycles.  Using the fact that only the real values of the fields enter the energy--momentum tensor, the time average of (\ref{energy-momentum-tensor}) is,
    \begin{widetext}
    \begin{equation}
        {\small\langle T^{\mu\nu}\rangle=
        \text{Re}\left(
                \begin{matrix}
			\frac{\epsilon_{0}}{4}\left[\left|\boldsymbol{E}_{\omega}\right|^{2}+c^2\left|\boldsymbol{B}_{\omega}\right|^{2}\right]&\frac{\epsilon_{0}c}{2}\left[\boldsymbol{E}_{\omega}\times\boldsymbol{B}_{\omega}^{\star}\right]\\[5pt]
			\frac{\epsilon_{0}c}{2}\left[\boldsymbol{E}_{\omega}\times\boldsymbol{B}_{\omega}^{\star
}\right]&
                    \langle T^{00}\rangle\mathbb{1}_{3}-\frac{\epsilon_{0}}{2}[\boldsymbol{E}_{\omega}
\otimes\boldsymbol{E}_{\omega}^{\star}+c^{2}\boldsymbol{B}_{\omega}\otimes\boldsymbol{B}_{\omega}^{\star}]
                \end{matrix}
            \right)}\\\label{averaged-energy-momentum-tensor}
    \end{equation}
    \end{widetext}
  For the specific case of a \textit{plane wave} reflecting off a surface, as shown \textit{e.g.} on the left hand side of the planar slab in fig. 1, the electric and magnetic fields are,
      \begin{align}
        \boldsymbol{E}_{\omega}(\mathbf{x})&=\sum_{\{q\}}\sum_{\{\pm\}}\hat{\boldsymbol{e}}_{q}^{ \text{\tiny{($\pm$)}}}\alpha_{q}^{\text{\tiny{($\pm$)}}}
        e^{i\boldsymbol{k}^{\text{\tiny{($\pm$)}}}\cdot\boldsymbol{x}}\nonumber\\
	\boldsymbol{B}_{\omega}(\mathbf{x})&=-\frac{1}{c}\sum_{\{q\}}\sum_{\{\pm\}}(-1)^{q}
        \hat {\boldsymbol{e}}_{\bar{q}}^{(\pm)}\alpha_{q}^{(\pm)}e^{i\boldsymbol{k}^{\text{\tiny{($\pm$)}}}\cdot\boldsymbol{x}}
        \label{electric-magnetic-fields}
\end{align}
    where the wave-vectors
    \begin{equation}
	\boldsymbol{k}^{\text{\tiny{($\pm$)}}}=\pm |k_{x}|\hat{\boldsymbol{x}}+k_{\parallel}\hat{\boldsymbol{k}}_{\parallel}.
    \end{equation}
    are decomposed as usual in terms of their normal and in--plane components.  For each wavevector there are two transverse polarization directions labelled by $q=1,2$ with  \(\bar{q}=(q\; \text{mod}\,2)+1\).  These polarization vectors are defined as,
    \begin{equation}
        \begin{tabular}{ll}
\(\hat{\boldsymbol{e}}_{1}^{(\pm)}=\hat{\boldsymbol{x}}\boldsymbol{\times}\hat{\boldsymbol{k}}_{\parallel
}=\frac{1}{k_{\parallel}}\left[k_{y}\hat{\boldsymbol{z}}-k_{z}\hat{\boldsymbol{y}}\right]
\)
& \quad$s$--polarization
        \end{tabular}\label{pol1}
    \end{equation}
 normal to the plane of incidence (TE polarization), and
        \begin{equation}
        \begin{tabular}{ll}
\(\hat{\boldsymbol{e}}_{2}^{(\pm)}=\hat{\boldsymbol{k}}^{(\pm)}\boldsymbol{\times}\hat{\boldsymbol{e}}_{1}=\zeta_{\parallel}
\hat{\boldsymbol{x}}\mp \zeta_{x}\hat{\boldsymbol{k}}_{\parallel}\;\)&$p$--polarization
        \end{tabular}\label{pol2}
    \end{equation}
      parallel to the plane of incidence (TM polarization), with the directional terms 
      \begin{equation}
      \zeta_{x}=c |k_{x}|/\omega=\cos\theta \quad \quad \zeta_{\parallel}=c k_{\parallel}/\omega=\sin\theta. 
      \label{angles}
      \end{equation}
Here \(\alpha_{1,2}^{(\pm)}\) are the electric field complex amplitudes of the right \((+)\) and left \((-)\) propagating waves, of either \(s\) (TE), or \(p\) (TM) polarization.  The summation in (\ref{electric-magnetic-fields}) is over the modes propagating to the right \((+)\) and to the left \((-)\), and over the two polarizations labelled by \(q\).  The geometrical details of the slab and the plane--wave configurations at the two interfaces on the left and the right are shown in figure \ref{figure-1}, where the origin of the $x$~axis is taken at the boundary between the medium and the free-space on the left-hand side of the slab and the $y$ and $z$ axis lie parallel to the interface.
\par
    We proceed to use (\ref{averaged-energy-momentum-tensor}) to calculate the radiation pressure experienced by the slab from a plane wave reflecting on the left.  Using the form of the fields given in (\ref{electric-magnetic-fields}), we can write down the general energy--momentum tensor \textit{outside} a planar surface in terms of the field amplitudes, \(\alpha_{1,2}^{(\pm)}\).  The time--averaged energy density from (\ref{averaged-energy-momentum-tensor}) is found to be,
    \begin{equation}
\langle\mathscr{E}_{\text{\tiny{F}}}\rangle=\frac{\epsilon_{0}}{2}\sum_{\{q\}}
\left[\sum_{\{\pm\}}\left|\alpha_{q}^{(\pm)}\right|^{2}+\zeta_{\parallel}^{2}A_{qq}
(x)\right],\label{general-energy-density}
    \end{equation}
    where the terms \(A_{qr}(x)=2\Re(\alpha^{(+)}_{q}\alpha^{(-)\star}_{r}e^{2i|k_{x}|x})\) are interpreted as being due to a stationary interference
pattern between incoming and outgoing waves.  The time--averaged momentum density is similarly,
    \begin{multline}
	c\langle\boldsymbol{\mathcal{P}}_{\text{\tiny{F}}}\rangle=\frac{\epsilon_{0}}{2}
	\sum_{\{q\}}\Bigg[\sum_{\{\pm\}}\left|\alpha_{q}^{(\pm)}\right|^{2}\hat{\boldsymbol{k}}^
	{(\pm)}\\
	+\zeta_{\parallel}\left(A_{qq}(x)\hat{\boldsymbol{k}}_{\parallel}+(-1)^{q}\zeta_{x}
	A_{q\bar{q}}(x)\hat{\boldsymbol{e}}_1\right)\Bigg],\label{general-momentum-density}
    \end{multline}
    where it is clear that there is a momentum flux parallel to the surface as well as normal to it.  The calculation of the \(3\times 3\) tensor, \(-\langle\boldsymbol{\sigma}_F\rangle\), is slightly more involved, yet noting that
    \[
	\hat{\boldsymbol{e}}_{1}\otimes\hat{\boldsymbol{e}}_{1}+\hat{\boldsymbol{e}}_{2}^{(\pm)}
	\otimes\hat{\boldsymbol{e}}_{2}^{(\pm)}-\mathbb{1}_{3}=-\hat{\boldsymbol{k}}^{(\pm)}
	\otimes\hat{\boldsymbol{k}}^{(\pm)},
    \]
    and,
    \begin{multline*}
	\hat{\boldsymbol{e}}_{1}\otimes\hat{\boldsymbol{e}}_{1}+\frac{1}{2}\left(\hat{\boldsymbol{e}
	}^{(-)}_{2}\otimes\hat{\boldsymbol{e}}_{2}^{(+)}+\hat{\boldsymbol{e}}_{2}^{(+)}
	\otimes\hat{\boldsymbol{e}}_{2}^{(-)}\right)-\zeta_\parallel^{2}\mathbb{1}_{3}\\
	=-\hat{\boldsymbol{k}}_{\parallel}\otimes\hat{\boldsymbol{k}}_{\parallel}+\zeta_{
	x}^{2}\hat{\boldsymbol{e}}_{1}\otimes\hat{\boldsymbol{e}}_{1},
    \end{multline*}
    one has,
    \begin{multline}
	-\langle\boldsymbol{\sigma}_{\text{\tiny{F}}}\rangle=\frac{\epsilon_{0}}{2}\sum_
	{\{q\}}\Bigg[\sum_{\{\pm\}}\left|\alpha_{q}^{(\pm)}\right|^{2}\hat{\boldsymbol{k}}^{
	(\pm)}\otimes\hat{\boldsymbol{k}}^{(\pm)}\\
	+A_{qq}(x)\left(\hat{\boldsymbol{k}}_{\parallel}\otimes\hat{\boldsymbol{k}}_{\parallel}
	-\zeta^{2}_{x}\hat{\boldsymbol{e}}_{1}\otimes\hat{\boldsymbol{e}}_{1}\right)\\
        +(-1)^{q}\zeta_{x}
	A_{q\bar{q}}(x)\left(\hat{\boldsymbol{k}}_{\parallel}\otimes\hat{\boldsymbol{e}}_{1}
	+\hat{\boldsymbol{e}}_{1}\otimes\hat{\boldsymbol{k}}_{\parallel}\right)\Bigg]\label{general-stress}
    \end{multline}
    Equations (\ref{general-energy-density}--\ref{general-stress}) determine the energy--momentum tensor for the case of a plane wave reflecting from the planar surface of a generic medium at rest, taking into account both polarizations.
    \par
    The time-averaged energy--momentum tensor given above allows us to determine the force exerted by the incident plane--wave on the medium, as well as the work done on it.  We consider fields in the rest frame of the medium and calculate the time-averaged four--momentum lost from the field per unit time in the space outside of the material.  The rate of change of the four--momentum density  \(\mathcal{P}_{\text{\tiny{F}}}^{\nu} = (\mathscr{E}_{\text{\tiny{F}}}/c, \boldsymbol{\mathcal{P}}_{\text{\tiny{F}}} \)) at any point outside of the medium is related to the spatial divergence of the energy--momentum tensor, \(T^{\mu\nu}\), via the vanishing four--divergence, \(\partial_{\mu}T^{\mu\nu}=0\), i.e,
    \begin{equation}
        \frac{1}{c}\frac{\partial T^{0\nu}}{\partial t}=\frac{\partial\mathcal{P}_{\text{\tiny{F}}}^{\nu}}{\partial t}=-\frac{\partial T^{i\nu}}{\partial x^{i}}.\label{force-density-relation}
    \end{equation}
    Integrating (\ref{force-density-relation}) over the volume of space outside of the material, \(V\), gives us the rate of change of the total four--momentum, \(P^{\nu}\), in free space,
    \[
        \frac{d P_{\text{\tiny{F}}}^{\nu}}{dt}=-\int_{V}\frac{\partial
T^{i\nu}}{\partial x^{i}}d^{3}\boldsymbol{x}=-\int_{\partial V}T^{i\nu}dS_{i},
    \]
    where the three--dimensional divergence theorem has been applied to obtain the final expression.  Due to the global conservation of four--momentum, whatever is lost in the free space is taken up by the region of space occupied by the medium,
    \[
        \frac{d P_{\text{\tiny{M}}}^{\nu}}{dt}=-\frac{d
P_{\text{\tiny{F}}}^{\nu}}{dt}=\int_{\partial V}T^{i\nu}dS_{i}
    \]
    Upon averaging over a time interval of many optical cycles this change of momentum is that taken up by the dielectric medium (the rate of change of electromagnetic momentum within this region of space averages to zero over such a time period).  Writing out the components explicitly using the notation introduced in (\ref{energy-momentum-tensor}), we  have,
        \begin{align}
	\left\langle\frac{d E_{\text{\tiny{M}}}}{dt}\right\rangle=c^{2}\int_{
	\partial V}\langle\boldsymbol{\mathcal{P}}_{\text{\tiny{F}}}\rangle\cdot
	d\boldsymbol{S}\nonumber\\
	\left\langle\frac{d\boldsymbol{P}_{\text{\tiny{M}}}}{dt}\right\rangle=-\int_{
	\partial V}\langle\boldsymbol{\sigma}_{\text{\tiny{F}}}\rangle\cdot
	d\boldsymbol{S},\label{four-force-equations}
    \end{align}
    where each surface element, \(d\boldsymbol{S}\) points \emph{into} the material (c.f. figure \ref{figure-1}).  In general, for an object of finite size in a plane wave field, there will be reflection from each surface of the object that will complicate the evaluation of the reflected fields and of the integrals in (\ref{four-force-equations}).  We assume throughout the case of a thin planar slab sufficiently extended over the \(y-z\) plane so that translational symmetry along this plane is preserved and the total force can be calculated integrating only over the left and right interfaces parallel to this plane.
    \par
    With these assumptions, and an application of (\ref{general-momentum-density}) and (\ref{general-stress}) to (\ref{four-force-equations}), the components of the time--averaged four--force experienced by such a medium in its rest frame are,
    \begin{align}
	\left\langle\frac{dE_{\text{\tiny{M}}}}{dt}\right\rangle&=\frac{
	A\epsilon_{0}c\zeta_{x}}{2}\sum_{\{q\}}\sum_{\{\pm\}}\pm\left(\left|\alpha_{qL}^{
	(\pm)}\right|^{2}-\left|\alpha_{qR}^{(\pm)}\right|^{2}\right)\nonumber\\
	\left\langle\frac{d\boldsymbol{P}_{\text{\tiny{M}}}}{dt}\right\rangle&=\frac{
	A\epsilon_{0}\zeta_{x}}{2}\sum_{\{q\}}\sum_{\{\pm\}}\pm\left(\left|\alpha_{qL}^{
	(\pm)}\right|^{2}-\left|\alpha_{qR}^{(\pm)}\right|^{2}\right)\hat{\boldsymbol{k}}^{
	(\pm)},\label{final-general-four-force}
    \end{align}
   where \(A\) is the integration area (the cross section of the beam).  The subscripts \(L\) and \(R\) indicate here the electric field amplitudes ($\alpha$) on the left (\(x<0\)), and right (\(x>d\)) of the slab, respectively (figure \ref{figure-1}). Hence, the time averaged four force on the slab is determined once the field amplitudes are known in vacuum.  For a given plane wave of frequency $\omega$ and wavevector $\boldsymbol{k}$,the two results in (\ref{final-general-four-force}) may be combined to yield the rather sound relation,
	\begin{equation}
		\left\langle\frac{d\boldsymbol{P}_{\text{\tiny{M}}}}{dt}\right\rangle
		=\frac{\boldsymbol{k}}{\omega}\left\langle\frac{dE_{\text{\tiny{M}}}}{dt}\right\rangle,
	\end{equation}
which holds true for either polarization.
%
%
\subsection{  \(4\times4\) Transfer and Scattering matrices.
\label{S-T-section}}
    \par
    In general, the relations between the amplitudes,  \(\alpha_{q\,L/R}^{(\pm)}\) in (\ref{final-general-four-force}) may be obtained by employing transfer or scattering matrix methods (e.g.~\cite{pendry1996,genet2003,artoni2005}), which we now use---along with the input fields---to determine the radiation pressure experienced by a general dielectric slab. The transfer matrix approach is ideally suited for the consideration of media inhomogeneous in one direction only (here taken along $x$) as the medium can be split up into as many layers as accuracy demands, and the transfer matrices of each of these layers applied one after the other to obtain the matrix associated with the whole slab. We will be using a \(4\times4\) transfer matrix theory as is commonly applied to anisotropic media (e.g.~\cite{schubert1996}).  This is because even if a medium is isotropic in its rest frame, when set in motion its symmetry is lowered.
%
%
\subsubsection{The transfer matrix}
\label{transfer-matrix-section}
    \par
    A typical \(4\times4\) transfer matrix, \(\boldsymbol{T}\), connects the field on the left of a slab (\(x<0\)) to the field on the
right (\(x>d\) see figure \ref{figure-1}) and is such that,
    \begin{equation}
\left(\begin{matrix}\boldsymbol{\alpha}_{1R}\\\boldsymbol{\alpha}_{2R}\end{matrix}\right)
        =\left(\begin{matrix}
                \boldsymbol{T}_{11}&\boldsymbol{T}_{12}\\
                \boldsymbol{T}_{21}&\boldsymbol{T}_{22}\\
        \end{matrix}\right)
    \left(\begin{matrix}\boldsymbol{\alpha}_{1L}\\\boldsymbol{\alpha}_{2L}\end{matrix}\right),\label{general-transfer-matrix}
    \end{equation}
    where the field amplitudes have been written in terms of,
    \[
        \boldsymbol{\alpha}_{q L/R}=\left(\begin{matrix}\alpha_{q
L/R}^{(+)}\\\alpha_{q L/R}^{(-)}\end{matrix}\right),
    \]
    and the transfer matrix has been compactly written as,
    \begin{equation}
        \boldsymbol{T}=\left(\begin{matrix}
                \boldsymbol{T}_{11}&\boldsymbol{T}_{12}\\
                \boldsymbol{T}_{21}&\boldsymbol{T}_{22}\\
        \end{matrix}\right)
        =
        \left(\begin{matrix}
            t_{11}&t_{12}&t_{13}&t_{14}\\
            t_{21}&t_{22}&t_{23}&t_{24}\\
            t_{31}&t_{32}&t_{33}&t_{34}\\
            t_{41}&t_{42}&t_{43}&t_{44}
        \end{matrix}\right)\label{expanded-transfer-matrix}
    \end{equation}
    with each sub--matrix, \(\boldsymbol{T}_{qr}\) relating amplitudes with polarization \(r\) on the left hand side of the medium, to amplitudes with polarization \(q\) on the right hand side. 
\par
      The elements of \textbf{\textit{T}} in (\ref{expanded-transfer-matrix}) are not all independent but in general depend on specific symmetries that the medium may posses. Most importantly, symmetries that hold in the medium rest frame may be lifted when the medium is set in motion---three specific cases of this are shown below.
\par
When the plane of incidence is a \textit{mirror symmetry plane}, then $s$ (TE) and $p$ (TM) polarizations do not mix, \textit{i.e.} are decoupled and 
the off diagonal sub--matrices of (\ref{expanded-transfer-matrix}) will vanish, \(\boldsymbol{T}_{12}=\boldsymbol{T}_{21}=0\), reducing  the transfer matrix in (\ref{general-transfer-matrix}) to the direct sum:
    \[
	\boldsymbol{T}=\boldsymbol{T}_{11}\oplus\boldsymbol{T}_{22}=\left(\begin{matrix}\boldsymbol{T}_{11}&\mathbf{0}\\
        \boldsymbol{0}&\boldsymbol{T}_{22}\end{matrix}\right).
    \]
This is \textit{e.g.}, the usual case for an isotropic medium at rest, for which a \(2\times2\) transfer matrix can instead be used, but no longer applies in general when the medium is set into motion. The plane of incidence remains in fact a plane of mirror symmetry for a medium moving along $x$, but not for a medium moving along $y$ (unless $k_z=0$) in which case the polarizations mix with non vanishing off diagonal elements in  (\ref{expanded-transfer-matrix}), as will be discussed below.  Meanwhile for a magneto--electric medium at rest, the constitutive relations take the form \(\boldsymbol{D}=\boldsymbol{\epsilon}\boldsymbol{\cdot}\boldsymbol{E}+\boldsymbol{\eta}\boldsymbol{\cdot}\boldsymbol{B}\)  and \(\boldsymbol{H}=\boldsymbol{\mu}^{-1}\boldsymbol{\cdot}\boldsymbol{B}+\boldsymbol{\zeta}\boldsymbol{\cdot}\boldsymbol{E}\)~\cite{volume8,lindell1994,lakhtakia1994} and, in general, all entries in (\ref{expanded-transfer-matrix}) are non--zero.  We may therefore refer to the polarization mixing induced by motion as to a ``\textit{magnetoelectric effect}''~\cite{volume8,leonhardt2010a}.
\par
The \textit{time reversal} for a plane wave is obtained by changing  $\boldsymbol{k}_{\parallel}$ into $-\boldsymbol{k}_{\parallel}$ and $\alpha_{q}^{(\pm)}$ into $(\alpha_{q}^{(\mp)})^{\star}$. When time reversal symmetry applies, e.g. for a lossless non-magnetized medium at rest, the transfer matrix should be the same when waves are interchanged in this way, i.e. left--going waves viewed with time running in reverse should be reflected and transmitted as if they were right--going incident waves with time running forwards,  provided that changing the sign of $k_y$ and $k_z$ is also immaterial as for an isotropic medium. Performing such a transformation on the transfer matrix leads to,
    \begin{align}
        \left(\begin{matrix}
                \boldsymbol{\sigma}_{x}\boldsymbol{T}_{11}^{\star}\boldsymbol{\sigma}_{x}&\boldsymbol{\sigma}_{x}\boldsymbol{T}_{12}^{\star}\boldsymbol{\sigma}_{x}\\
                \boldsymbol{\sigma}_{x}\boldsymbol{T}_{21}^{\star}\boldsymbol{\sigma}_{x}&\boldsymbol{\sigma}_{x}\boldsymbol{T}_{22}^{\star}\boldsymbol{\sigma}_{x}\\
        \end{matrix}\right)_{-\hat{\boldsymbol{k}}_{\parallel}}
    	=\left(\begin{matrix}
                \boldsymbol{T}_{11}&\boldsymbol{T}_{12}\\
                \boldsymbol{T}_{21}&\boldsymbol{T}_{22}\\
        \end{matrix}\right)_{\hat{\boldsymbol{k}}_{\parallel}}
	\label{tr-condition}
    \end{align}
where \(\boldsymbol{\sigma}_{x}\) is the usual Pauli matrix (\(\boldsymbol{\sigma}_{x}^{2}=\mathbb{1}_{2}\)). Therefore, (\ref{tr-condition}) shows that time reversal symmetry requires that the sub--matrices of \(\boldsymbol{T}\) transform as, \(\boldsymbol{T}_{ij}(\hat{\boldsymbol{k}}_{\parallel})=\boldsymbol{\sigma}_{x}\boldsymbol{T}_{ij}^{\star}(-\hat{\boldsymbol{k}}_{\parallel})\boldsymbol{\sigma}_{x}\).  For example,
  \[
    	\textbf{\textit{T}}_{11}=\left(\begin{matrix}t_{11}&t_{12}\\t_{21}&t_{22}\end{matrix}\right)_{\hat{\boldsymbol{k}}_{\parallel}}=\left(\begin{matrix}t_{22}^{\star}&t_{21}^{\star}\\t_{12}^{\star}&t_{11}^{\star}\end{matrix}\right)_{-\hat{\boldsymbol{k}}_{\parallel}}
    \]
For a moving medium, time reversal symmetry does not apply as it would require the velocity of the medium to change sign.
\par
If the median plane of a slab is a plane of mirror symmetry,  an input of a given amplitude produces the same reflected and transmitted amplitudes whether incident from the right or the left of the slab, and we refer to this as \textit{left--right symmetry}.  Noting that \(\boldsymbol{T}\) transfers amplitudes on the left of the slab to the right, and that \(\boldsymbol{T}^{-1}\) transfers amplitudes from right to left, this condition is equivalent to,
    \begin{equation}
    	\left(\begin{matrix}\boldsymbol{T}_{11}&\boldsymbol{T}_{12}\\\boldsymbol{T}_{21}&\boldsymbol{T}_{22}\end{matrix}\right)=
\left(\begin{matrix}\boldsymbol{\sigma}_{x}&\boldsymbol{0}\\\boldsymbol{0}&-\boldsymbol{\sigma}_{x}\end{matrix}\right)
\left(\begin{matrix}\tilde{\boldsymbol{T}}_{11}&\tilde{\boldsymbol{T}}_{12}\\\tilde{\boldsymbol{T}}_{21}&\tilde{\boldsymbol{T}}_{22}\end{matrix}\right)
\left(\begin{matrix}\boldsymbol{\sigma}_{x}&\boldsymbol{0}\\\boldsymbol{0}&-\boldsymbol{\sigma}_{x}\end{matrix}\right)
\label{reciprocity}
    \end{equation}
    where the \(\tilde{\boldsymbol{T}}_{ij}\) are the sub--matrices of the inverse, \(\boldsymbol{T}^{-1}\).  The \(\boldsymbol{\sigma}_{x}\) matrices are present to interchange left and right propagating waves, as we should in examining the mirror symmetric situation.  The reason for the minus sign in one of these matrices (\(-\boldsymbol{\sigma}_{x}\)) in (\ref{reciprocity}) is more subtle: it is present due to the definition of p--polarization (\ref{pol2}) (see figure \ref{figure-1}).  If the polarization of the incoming p--polarized wave on the right hand side of the slab points away from the surface then that on the left hand side points into the surface.  Therefore in examining the mirror symmetric situation we should multiply the p--polarized amplitudes by \(-1\).
    \par
    In general, the conditions for left--right symmetry arising from (\ref{reciprocity}) are complicated as they require the calculation of the inverse of a \(4\times4\) matrix.  However, in the case of media for which  \(\boldsymbol{T}=\boldsymbol{T}_{11}\oplus\boldsymbol{T}_{22}\), the condition becomes rather compact.  In this case it requires the elements of \(\boldsymbol{T}_{11}\) to obey,
\begin{equation}
    	 \boldsymbol{T}_{11}=\left(\begin{matrix}t_{11}&t_{12}\\t_{21}&t_{22}\end{matrix}\right)=\frac{1}{\det(\boldsymbol{T}_{11})}\left(\begin{matrix}t_{11}&-t_{21}\\-t_{12}&t_{22}\end{matrix}\right)\label{standard-reciprocity}
    \end{equation}
   with an identical relation also holding for the matrix \(\boldsymbol{T}_{22}\).  For this particular case, the \(\boldsymbol{T}_{ii}\) matrices must be antisymmetric and have a determinant equal to one (when such a medium is also time reversible, \(t_{22}=t_{11}^{\star}\), and \(t_{21}=t_{12}^{\star}=-t_{12}\)).  For  radiation passing from vacuum into a such a medium and then back into vacuum (as considered here), this determinant will always equal unity.  We finally note that when a medium is moving along $\hat{\boldsymbol{x}}$, left--right symmetry is lifted, while it still applies if the motion is perpendicular to $\hat{\boldsymbol{x}}$.
%
%
\subsubsection{The scattering matrix}
    \par
    Although undoubtedly useful for calculations, there is a slight awkwardness in the application of the transfer matrix to an experimental situation.  In general the known quantities are the `input' field amplitudes, \(\alpha_{qL}^{(+)}\) and \(\alpha_{qR}^{(-)}\), rather than either of the quantities, 
\(\boldsymbol{\alpha}_{q L/R}\) (see figure \ref{figure-1}).  A scattering matrix, \(\boldsymbol{S}\), rather than transfer matrix relates input and output,
    \begin{equation}
	\left(\begin{matrix}\boldsymbol{\alpha}_{1\,\text{\tiny{(OUT)}}}\\\boldsymbol{
	\alpha}_{2\,\text{\tiny{(OUT)}}}\end{matrix}\right)
        =\left(\begin{matrix}
                \boldsymbol{S}_{11}&\boldsymbol{S}_{12}\\
                \boldsymbol{S}_{21}&\boldsymbol{S}_{22}\\
        \end{matrix}\right)
	\left(\begin{matrix}\boldsymbol{\alpha}_{1\,\text{\tiny{(IN)}}}\\\boldsymbol{
	\alpha}_{2\,\text{\tiny{(IN)}}}\end{matrix}\right),\label{general-scattering-matrix}
    \end{equation}
    where,
    \begin{equation}
	\boldsymbol{S}=\left(\begin{matrix}\boldsymbol{S}_{11}&\boldsymbol{S}_{12}\\\boldsymbol{S}_{21}
	&\boldsymbol{S}_{22}\end{matrix}\right)
        =\left(\begin{matrix}
	\mathcal{T}_{11}&\bar{\mathcal{R}}_{11}&\mathcal{T}_{12}&\bar{\mathcal{R}}_{12}
	\\
	\mathcal{R}_{11}&\bar{\mathcal{T}}_{11}&\mathcal{R}_{12}&\bar{\mathcal{T}}_{12}
	\\
	\mathcal{T}_{21}&\bar{\mathcal{R}}_{21}&\mathcal{T}_{22}&\bar{\mathcal{R}}_{22}
	\\
	\mathcal{R}_{21}&\bar{\mathcal{T}}_{21}&\mathcal{R}_{22}&\bar{\mathcal{T}}_{22}
	\\	
        \end{matrix}\right),\label{expanded-scattering-matrix}
    \end{equation}
    with $\mathcal{R}_{qr} (\bar{\mathcal{R}}_{qr})$, $\mathcal{T}_{qr} (\bar{\mathcal{T}}_{qr})$ representing reflection and transmission coefficients for  \(q\)-polarized waves, given an \(r\)--polarized wave impinging from the left (right), and where,
    \begin{equation}
	\boldsymbol{\alpha}_{q\,\text{\tiny{(IN)}}}=\left(\begin{matrix}\alpha_{qL}^
	{(+)}\\\alpha_{qR}^{(-)}\end{matrix}\right),
	\label{input-amplitudes}
    \end{equation}
    and,
    \begin{equation}
	\boldsymbol{\alpha}_{q\,\text{\tiny{(OUT)}}}=\left(\begin{matrix}\alpha_{qR}^{(+)}\\\alpha_{qL}^
	{(-)}\end{matrix}\right).\label{output-amplitudes}
    \end{equation}
    Direct evaluation of the scattering matrix for a generic slab is in general rather involved.  One would then proceed, first finding the transfer matrix for the whole slab, and then converting this into a scattering matrix for the calculation of the force, via (\ref{final-general-four-force}).
    \par
    The relationship between the elements of the scattering matrix (\ref{expanded-scattering-matrix}) and the transfer matrix can be found through expanding the matrix equation (\ref{general-transfer-matrix}) in terms of the \(16\) elements of (\ref{expanded-transfer-matrix}), and writing the output field amplitudes, (\ref{output-amplitudes}), in terms of the input amplitudes and the \(t_{ij}\).  This gives,
\[
	\boldsymbol{S}=\left(\boldsymbol{\pi}^{(-)}+\boldsymbol{\pi}^{(+)}\boldsymbol{\cdot}\boldsymbol{T}\right)\boldsymbol{\cdot}\left(\boldsymbol{\pi}^{(+)}+\boldsymbol{\pi}^{(-)}\boldsymbol{\cdot}\boldsymbol{T}\right)^{-1}
\]
where we have introduced two matrices, \(\boldsymbol{\pi}^{\text{\tiny{($+$)}}}=\text{diag}(1,0,1,0)\) and \(\boldsymbol{\pi}^{\text{\tiny{($-$)}}}=\text{diag}(0,1,0,1)\).  Explicitly evaluating the components of the \(\boldsymbol{S}\) matrix we have
    \begin{widetext}
    \begin{equation}
        \boldsymbol{S}=
        \frac{1}{(t_{22}t_{44}-t_{24}t_{42})}
        \left(\begin{matrix}
            M_{33} &(t_{12}t_{44}-t_{14}t_{42})    &   -M_{31} & (t_{22}t_{14}-t_{12}t_{24})\\
            (t_{24}t_{41}-t_{44}t_{21}) &   t_{44} &   (t_{24}t_{43}-t_{44}t_{23}) &   -t_{24}\\
            -M_{13} &(t_{32}t_{44}-t_{34}t_{42})    &   M_{11} &(t_{22}t_{34}-t_{32}t_{24})\\
            (t_{42}t_{21}-t_{22}t_{41}) &   -t_{42} &   (t_{42}t_{23}-t_{22}t_{43}) &   t_{22}
        \end{matrix}\right),\label{scattering-transfer-relationship}
    \end{equation}
    \end{widetext}
    where \(M_{ij}\) is the minor \((i,j)\) of \(\boldsymbol{T}\), e.g. 
    \[
	M_{13}=\left|\begin{matrix}t_{21}&t_{22}&t_{24}\\
		t_{31}&t_{32}&t_{34}\\
		t_{41}&t_{42}&t_{44}\end{matrix}\right|.
    \]
    Note that, as expected, when the magnetoelectric effect is not present, the scattering matrix also reduces to a direct sum, \(\boldsymbol{S}=\boldsymbol{S}_{11}\oplus\boldsymbol{S}_{22}\), with the comparatively simple form,
    \[
        \boldsymbol{S}_{11}=\frac{1}{t_{22}}
        \left(\begin{matrix}
       1&t_{12}\\
        -t_{21}&1
        \end{matrix}\right),\;
        \boldsymbol{S}_{22}=\frac{1}{t_{44}}
        \left(\begin{matrix}
        1&t_{34}\\
        -t_{43}&1
        \end{matrix}\right)
    \]
    where we have assumed that \( \text{det}\left(\boldsymbol{T}_{11}\right)= \text{det}\left(\boldsymbol{T}_{22}\right)=1\).  From (\ref{scattering-transfer-relationship}) it is clear that in the general case there is quite an intricate relationship between the scattering and transfer matrices.  We also note that in the case of systems with a resonance in reflection or transmission (e.g. a Fabry--Perot cavity), the resonant frequency is generally determined by \(t_{22}t_{44}=t_{24}t_{42}\), rather than the usual separate conditions for the two polarizations, \(t_{22}=0\) and \(t_{44}=0\).  In addition, if the medium is left--right symmetric, one has \(t_{21}=-t_{12}\) and \(t_{43}=-t_{34}\), so that; \(\mathcal{R}_{11}=\bar{\mathcal{R}}_{11}=t_{12}/t_{22}\); \(\mathcal{T}_{11}=\bar{\mathcal{T}}_{11}=1/t_{22}\); \(\mathcal{R}_{22}=\bar{\mathcal{R}}_{22}=t_{34}/t_{44}\); and, \(\mathcal{T}_{22}=\bar{\mathcal{T}}_{22}=1/t_{44}\).  Due to the restrictions of the previous section (\(t_{12}=-t_{12}^{\star}\)), time reversibility has the further consequence that, \(\mathcal{R}^{\star}_{qq}\mathcal{T}_{qq}+\mathcal{T}_{qq}^{\star}\mathcal{R}_{qq}=0\).
    \par
    Finally, in terms of the scattering matrix elements defined in (\ref{general-scattering-matrix}) the four--force (\ref{final-general-four-force}) can be rewritten as,
    \begin{align}
	\left\langle\frac{dE_{\text{\tiny{M}}}}{dt}\right\rangle&=\frac{
	A\epsilon_{0}\zeta_{x}c}{2}\boldsymbol{\alpha}_{\text{\tiny{(IN)}}}^{\dagger}\boldsymbol{\cdot}
	\left(\mathbb{1}_{4}-\boldsymbol{S}^{\dagger}\boldsymbol{S}\right)\boldsymbol{\cdot}\boldsymbol{\alpha}_{
	\text{\tiny{(IN)}}}
	\nonumber\\
	\left\langle\frac{d\boldsymbol{P}_{\text{\tiny{M}}}}{dt}\right\rangle&=\frac{
	A\epsilon_{0}\zeta_{x}}{2}\boldsymbol{\alpha}_{\text{\tiny{(IN)}}}^{\dagger}
	\boldsymbol{\cdot}\left(\begin{matrix}\left(\boldsymbol{R}-\boldsymbol{S}^{\dagger}\boldsymbol{R}\boldsymbol{S}
	\right)&\zeta_{x}\hat{\boldsymbol{x}}\\
	\left(\mathbb{1}_{4}-\boldsymbol{S}^{\dagger}\boldsymbol{S}\right)&\zeta_{\parallel}\hat
	{\boldsymbol{k}}_{\parallel}\end{matrix}\right)\boldsymbol{\cdot}\boldsymbol{\alpha}_{\text{\tiny{(IN)
	}}}
	\label{four-force-scattering-matrix}
    \end{align}
    where the matrices in the second equation operate on the input amplitudes outside of the vector, and \(\boldsymbol{R}=\boldsymbol{\sigma}_{z}\oplus\boldsymbol{\sigma}_{z}\), with \(\boldsymbol{\sigma}_{z}\) the usual Pauli matrix.  Hence, once we have determined the transfer matrix for any given planar slab we have, via (\ref{scattering-transfer-relationship}--\ref{four-force-scattering-matrix}), thereby determined the four--force, for any given input field.  In (\ref{four-force-scattering-matrix}) only the input field amplitudes enter, which is the main advantage of (\ref{four-force-scattering-matrix}) over (\ref{final-general-four-force}).  When the slab is lossless, the scattering matrix is unitary, \(\boldsymbol{S}^{\dagger}\boldsymbol{S}=\mathbb{1}_{4}\), and any lateral force on the slab will vanish.
    \par
    For the simplest case of incidence from the left (\(\alpha_{q\;\text{\tiny{R}}}^{\text{\tiny{(-)}}}=0\)), onto a reciprocal, non--magneto--electric medium, equation (\ref{four-force-scattering-matrix}) can be expanded, using the notation of (\ref{expanded-scattering-matrix}), to give the usual result (e.g. within~\cite{loudon2002} these expressions appear for a single polarization in the force experienced by a transparent dielectric slab),
    \begin{align}
	{\small\left\langle\frac{dE_{\text{\tiny{M}}}}{dt}\right\rangle}&{
	\small=\frac{A\epsilon_{0}c\zeta_{x}}{2}\sum_{\{q\}}\left|\alpha_{q\text{\tiny{L}}}^
	{\text{\tiny{($+$)}}}\right|^{2}\left(1-\left|\mathcal{R}_{qq}\right|^{2}
	-\left|\mathcal{T}_{qq}\right|^{2}\right)}\nonumber\\
	{\small\left\langle\frac{d\boldsymbol{P}_{\text{\tiny{M}}}}{dt}\right\rangle}&{
	\small=\frac{A\epsilon_{0}\zeta_{x}}{2}\sum_{\{q\}}\left|\alpha_{q\text{\tiny{L}}}^{
	\text{\tiny{($+$)}}}\right|^{2}\left(\begin{matrix}[1+\left|\mathcal{R}_{qq}
	\right|^{2}-\left|\mathcal{T}_{qq}\right|^{2}]&\zeta_{x}\hat{\boldsymbol{x}}\\
    	[1-\left|\mathcal{R}_{qq}\right|^{2}-\left|\mathcal{T}_{qq}\right|^{2}]
	&\zeta_{\parallel}\hat{\boldsymbol{k}}_{\parallel}\end{matrix}\right)}\label{reciprocal-force},
    \end{align}
    It is clear that the lateral force is simply proportional to the absorption \(\left|\mathcal{A}^{}_{qq}\right|^{2}=(1-\left|\mathcal{R}^{}_{qq}\right|^{2}-\left|\mathcal{T}^{}_{qq}\right|^{2})\), while the normal force contains, besides the contribution of the absorption, a term proportional to twice the reflectivity, i.e. \(2\left|\mathcal{R}^{}_{qq}\right|^{2}+\left|\mathcal{A}^{}_{qq}\right|^{2}=(1+\left|\mathcal{R}^{}_{qq}\right|^{2}-\left|\mathcal{T}^{}_{qq}\right|^{2})\), as reflection entails the exchange of twice the momentum in the direction normal to the surface.
    \par
    To conclude this section we note that in the above we have not yet assumed that the dielectric is in motion.  We shall find in the following section that for certain regimes, the motion of the medium induces a coupling between the s and p polarizations.  However, the above conclusions hold for any media that couple polarizations, moving or otherwise.
%
%
\section{Moving media and radiation pressure\label{mmsec}}
    \par
   In the following we use the formalism of section \ref{tmsec} to study radiation pressure forces on a moving slab, and we will specifically examine two regimes of motion, namely, when the slab moves parallel (A.) and perpendicular (B.) to the interface.
%
%
    \subsection{Motion in the \(y\)--\(z\) plane\label{Vy-section}}
    \par
\begin{figure}[hc]
	\includegraphics[width=6.5cm]{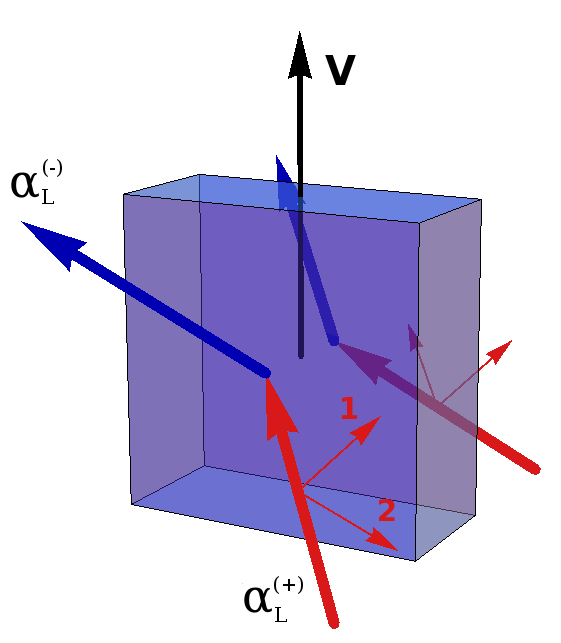}
	\caption{For certain angles of incidence, a laterally moving medium mixes polarizations in transmission and reflection.  For instance, an s--polarized input will generate p--polarized reflected and transmitted amplitudes so long as \(k_{z}\neq0\).  In all cases the Doppler shift makes the reflection and transmission coefficients dependent upon \(k_{y}\) in such a way that, for example \(\mathcal{R}_{qr}(k_{y})\neq \mathcal{R}_{qr}(-k_{y})\).\label{figure-2}}
    \end{figure}

    When an incident plane wave of a frequency \(\omega\) reflects from a surface that moves in the plane orthogonal to the surface normal, the reflected plane wave also has frequency \(\omega\), and the field is monochromatic in both rest frame and lab frame.  Therefore we may apply the result (\ref{four-force-scattering-matrix}) directly.  Without loss of generality, we take the motion to be along \(\hat{\boldsymbol{y}}\) to be representative of the general motion in the  \(y\)--\(z\)  plane (see figure \ref{figure-2}).  
%
%
             \begin{figure}[hc]
 	\includegraphics[width=8.3cm]{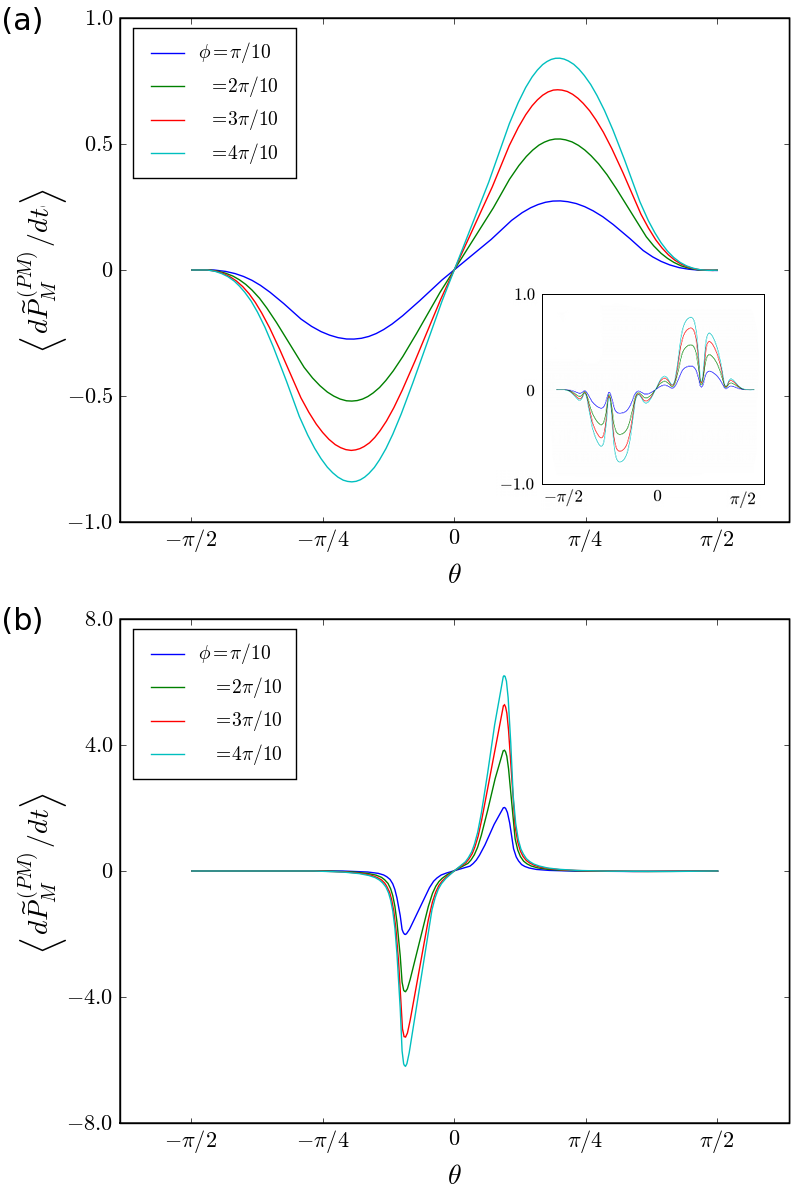}
	\caption{Contribution to the normal force (\ref{first-order-normal-force}) from the mixing of polarizations due to the motion of the medium.  Here we plot the quantity, \(\langle d\tilde{P}^{\text{\tiny{(PM)}}}_{M}/dt\rangle=\zeta_{x}^{2}\sum_{q}2(-1)^{\bar{q}}\eta(|\mathcal{R}_{qq}|^{2}-|\mathcal{T}_{qq}|^{2})\).  This represents the contribution to the normal force due to polarization mixing when \(\alpha_{1L}^{(+)}=\alpha_{2L}^{(+)}\), normalized in units of the incident momentum flux times \(V_{y}/c\).  In figure (a) we plot \(\langle dP^{\text{\tiny{(PM)}}}_{M}/dt\rangle\) as a function of \(\theta\) for various \(\phi\), where \(\boldsymbol{k}^{(+)}=(\omega/c)[\cos(\theta)\hat{\boldsymbol{x}}+\sin(\theta)(\cos(\phi)\hat{\boldsymbol{y}}+\sin(\phi)\hat{\boldsymbol{z}})]\).  In the main plot the radiation is incident onto a slab of thickness \(10c/\omega\) with \(\epsilon=6.0+0.01i\) \& \(\mu=1.0\).  The inset shows the case when the thickness is \(40c/\omega\).  Figure (b) illustrates the increase in magnitude of \(\langle dP^{\text{\tiny{(PM)}}}_{M}/dt\rangle\) when the refractive index drops below unity:  \(\epsilon=0.1+0.01i\) \& \(\mu=1.0\), where the medium becomes sensitive to the difference between s and p polarization at small angles of incidence (in this regime \(\eta=\cot(\theta)\sin(\phi)\) is large).  Within both plots we used the reflection and transmission coefficients for a slab of dielectric of arbitrary thickness as can be found in~\cite{born2009}\label{figure-2a}}
 \end{figure}

    \par
    In the \textit{rest frame} of the medium, we apply (\ref{four-force-scattering-matrix}),
    \begin{align}
	\left\langle\frac{dE_{\text{\tiny{M}}}^{\prime}}{dt^{\prime}}\right\rangle&=\frac{
	A^{\prime}\epsilon_{0}\zeta^{\prime}_{x}c}{2}\boldsymbol{\alpha}_{\text{\tiny{(IN)}}}^{\prime\dagger}
	\boldsymbol{\cdot}\left(\mathbb{1}_{4}-\boldsymbol{S}^{\prime\dagger}\boldsymbol{S}^{\prime}\right)\boldsymbol{\cdot}\boldsymbol{\alpha}_{\text{\tiny{(IN)}}}^{\prime}\nonumber\\
	\left\langle\frac{d\boldsymbol{P}_{\text{\tiny{M}}}^{\prime}}{dt^{\prime}}\right\rangle&=\frac{
	A^{\prime}\epsilon_{0}\zeta_{x}^{\prime}}{2}\boldsymbol{\alpha}_{\text{\tiny{(IN)}}}^{\prime\dagger}
	\boldsymbol{\cdot}\left(\begin{matrix}\left(\boldsymbol{R}-\boldsymbol{S}^{\prime\dagger}\boldsymbol{R}\boldsymbol{S}^{\prime}
	\right)&\zeta^{\prime}_{x}\hat{\boldsymbol{x}}\\
	\left(\mathbb{1}_{4}-\boldsymbol{S}^{\prime\dagger}\boldsymbol{S}^{\prime}\right)&\zeta^{\prime}_{\parallel}\hat
	{\boldsymbol{k}}^{\prime}_{\parallel}\end{matrix}\right)\boldsymbol{\cdot}\boldsymbol{\alpha}^{\prime}_{\text{\tiny{(IN)}}}\label{lateral-force}
    \end{align}
    where the primed  matrices, \(\mathbf{S}^{\prime}\), contain reflection and transmission coefficients evaluated at the rest frame frequency and wave--vector.
    \par
   The field amplitudes with different polarizations transform between laterally moving frames as follows (See appendix \ref{appendix-lateral-motion}),
    \begin{equation}
    	\boldsymbol{\alpha}^{\prime}_{\text{\tiny{(IN)}}}=\left(\frac{\omega^{\prime}}{\omega}\right)\boldsymbol{M}\boldsymbol{\cdot}\boldsymbol{\alpha}_{\text{\tiny{(IN)}}},\label{field-transform}
    \end{equation}
    where,
    \[
	\boldsymbol{M}=\frac{1}{\sqrt{1+\frac{V_{y}^{2}\eta^{2}}{c^{2}}}}\left(\begin{matrix}\mathbb{1}_{2}&\frac{V_{y}\eta}{c}\boldsymbol{\sigma}_{z}\\-\frac{V_{y}\eta}{c}\boldsymbol{\sigma}_{z}&\mathbb{1}_{2}\end{matrix}\right)
    \]
 has off-diagonal terms directly proportional to the mixing parameter 
    \begin{equation}
    \eta=\frac{|k_{x}|k_{z}}{k_{\parallel}^{2}-\omega V_{y}k_{y}/c^{2}}.
    \end{equation}
    The matrix, \(\boldsymbol{M}\) is unitary, and the magnitude of the field intensity, \(\boldsymbol{\alpha}^{\dagger}\boldsymbol{\alpha}\), is therefore only modified by the Doppler shift through the factor, \((\omega^{\prime}/\omega)^{2}\).  Using (\ref{field-transform}), and Lorentz transforming (\ref{lateral-force}) to the lab frame, we arrive at the four--force experienced by a material in lateral motion,
      \begin{align}
        \left\langle\frac{dE_{\text{\tiny{M}}}}{dt}\right\rangle&=\frac{
        A\epsilon_{0}c\zeta_{x}}{2}\boldsymbol{\alpha}_{\text{\tiny{(IN)}}}^{\dagger}
        \left(\mathbb{1}_{4}-\tilde{\boldsymbol{S}}^{\dagger}\tilde{\boldsymbol{S}}
        \right)\boldsymbol{\alpha}_{\text{\tiny{(IN)}}}\nonumber\\
        \left\langle\frac{d\boldsymbol{P}_{\text{\tiny{M}}}}{dt}\right\rangle&=\frac{
        A\epsilon_{0}\zeta_{x}}{2}\boldsymbol{\alpha}_{\text{\tiny{(IN)}}}^{\dagger}\left(\begin{matrix}
        \left(\boldsymbol{R}-\tilde{\boldsymbol{S}}^{\dagger}\boldsymbol{R}\tilde{\boldsymbol{S}}\right)&\zeta_{x}\hat{\boldsymbol{x}}\\
        \left(\mathbb{1}_{4}-\tilde{\boldsymbol{S}}^{\dagger}\tilde{\boldsymbol{S}}\right)&\zeta_{\parallel}\hat{\boldsymbol{k}}_{\parallel}\end{matrix}\right)
        \boldsymbol{\alpha}_{\text{\tiny{(IN)}}}\label{lateral-result}
      \end{align}
      where, \(A=A^{\prime}/\gamma\), and,
      \begin{equation}
          \tilde{\boldsymbol{S}}=\boldsymbol{M}^{\dagger}\boldsymbol{S}^{\prime}\boldsymbol{M}\label{unitary-transformed-S}
      \end{equation}
\par     
The significance of (\ref{lateral-result}) is that the four--force experienced by a laterally moving medium can be completely described by a unitary transformation of the rest frame scattering matrix $\boldsymbol{S}^{\prime}$.  The physics of radiation pressure on laterally moving media is entirely wrapped up in the dependence of the elements of \(\boldsymbol{S}^{\prime}\) on the Lorentz transformed frequency and wave--vector, generally referred here as to Doppler shift,  and in the magnetoelectric effect induced by the unitary transformation, $\boldsymbol{M}$.  Notice that when the angle of incidence is such that either \(k_{x}\) or \(k_{z}\) equal zero, then \(\eta=0\), so that \(\tilde{\boldsymbol{S}}=\boldsymbol{S}^{\prime}\), and the only effect distinguishing the laterally moving medium from a stationary one is related to the Doppler shift~\footnote{This is true with the exception of the case of normal incidence where polarization mixing is delicate due to the degeneracy of \(s\) and \(p\) polarizations.  Such delicate behaviour, demonstrated within the fact that \(\eta\) can take large values close to normal incidence is usually but not always without physical significance.  For typical dielectric media, the response is degenerate to the two types of polarization for small angles of incidence.  However, see figure~\ref{figure-2a}b.}.
      \par
      Using the notation of (\ref{general-scattering-matrix}), the unitary transformation of the scattering matrix given in (\ref{unitary-transformed-S}) for a non magneto-electric medium where  \(\boldsymbol{S}_{12}^{\prime}=\boldsymbol{S}_{21}^{\prime}=\boldsymbol{0}\) is,
   \begin{equation}
          \tilde{\boldsymbol{S}}=\left(\begin{matrix}
                      \frac{\boldsymbol{S}^{\prime}_{11}+\frac{\eta^{2}V_{y}^{2}}{c^{2}}\boldsymbol{\sigma}_{z}\boldsymbol{S}^{\prime}_{22}\boldsymbol{\sigma}_{z}}{1+\frac{V_{y}^{2}\eta^{2}}{c^{2}}}&\frac{V_{y}\eta}{c}\frac{\left(\boldsymbol{S}^{\prime}_{11}\boldsymbol{\sigma}_{z}-\boldsymbol{\sigma}_{z}\boldsymbol{S}^{\prime}_{22}\right)}{1+\frac{V_{y}^{2}\eta^{2}}{c^{2}}}\\
                                 \frac{V_{y}\eta}{c}\frac{\left(\boldsymbol{\sigma}_{z}\boldsymbol{S}^{\prime}_{11}-\boldsymbol{S}^{\prime}_{22}\boldsymbol{\sigma}_{z}\right)}{1+\frac{V_{y}^{2}\eta^{2}}{c^{2}}}&\frac{\boldsymbol{S}^{\prime}_{22}+\frac{\eta^{2}V_{y}^{2}}{c^{2}}\boldsymbol{\sigma}_{z}\boldsymbol{S}^{\prime}_{11}\boldsymbol{\sigma}_{z}}{1+\frac{V_{y}^{2}\eta^{2}}{c^{2}}}
                                   \end{matrix}\right)\label{transformed-scattering-matrix}
    \end{equation}\\
    There are several interesting features of (\ref{transformed-scattering-matrix}).  Firstly the diagonal sub--matrices, \(\tilde{\boldsymbol{S}}_{11}\) \& \(\tilde{\boldsymbol{S}}_{22}\) are only modified by the transformation (\ref{unitary-transformed-S}) via terms that are second order in the velocity, and such terms are typically negligible.  Secondly, the off diagonal terms linear in the velocity, \(\tilde{\boldsymbol{S}}_{12}\) \& \(\tilde{\boldsymbol{S}}_{21}\), are  related by \(\tilde{\boldsymbol{S}}_{21}=\boldsymbol{\sigma}_{z}\tilde{\boldsymbol{S}}_{12}\boldsymbol{\sigma}_{z}\).  Therefore, the mixed polarization reflection coefficients are related by a minus sign (\(\tilde{\mathcal{R}}_{21}=-\tilde{\mathcal{R}}_{12}\)), while the corresponding  transmission coefficents are equal (\(\tilde{\mathcal{T}}_{21}=\tilde{\mathcal{T}}_{12}\)).  Assuming a rest frame scattering matrix which is left--right symmetric, an examination of the off diagonal matrices in (\ref{transformed-scattering-matrix}) shows that \(\tilde{\mathcal{R}}_{21}=-\bar{\tilde{\mathcal{R}}}_{21}\), \(\tilde{\mathcal{R}}_{12}=-\bar{\tilde{\mathcal{R}}}_{12}\), \(\tilde{\mathcal{T}}_{21}=-\bar{\tilde{\mathcal{T}}}_{21}\), and \(\tilde{\mathcal{T}}_{12}=-\bar{\tilde{\mathcal{T}}}_{12}\), which means that left--right symmetry still holds in this case as expected (recall the minus sign within the definition of the polarization shown in figure \ref{figure-1}).    
\par
     With the help of (\ref{transformed-scattering-matrix}) we may now write out the four--force (\ref{lateral-result}) for the simplest case of a plane wave incident on the slab from the left,
    \begin{widetext}
        \begin{align}
          \left\langle\frac{dE_{\text{\tiny{M}}}}{dt}\right\rangle&=\frac{
          A\epsilon_{0}c\zeta_{x}}{2\left(1+\frac{V_{y}^{2}\eta^{2}}{c^{2}}\right)}\sum_{\{q\}}\left(1-\left|\mathcal{R}^{\prime}_{qq}\right|^{2}-\left|\mathcal{T}^{\prime}_{qq}\right|^{2}\right)
          \left|\alpha_{\text{\tiny{$q L$}}}^{\text{\tiny{($+$)}}}+(-1)^{\bar{q}}\frac{V_{y}\eta}{c}\alpha_{\text{\tiny{$\bar{q} L$}}}^{\text{\tiny{($+$)}}}\right|^{2}\nonumber\\
          \left\langle\frac{d\boldsymbol{P}_{\text{\tiny{M}}}}{dt}\right\rangle&=\frac{
          A\epsilon_{0}\zeta_{x}}{2\left(1+\frac{V_{y}^{2}\eta^{2}}{c^{2}}\right)}\sum_{\{q\}}\left(\begin{matrix}
          \left(1+\left|\mathcal{R}^{\prime}_{qq}\right|^{2}-\left|\mathcal{T}^{\prime}_{qq}\right|^{2}\right)&\zeta_{x}\hat{\boldsymbol{x}}\\
          \left(1-\left|\mathcal{R}^{\prime}_{qq}\right|^{2}-\left|\mathcal{T}^{\prime}_{qq}\right|^{2}\right)&\zeta_{\parallel}\hat{\boldsymbol{k}}_{\parallel}\end{matrix}\right)
          \left|\alpha_{\text{\tiny{$q L$}}}^{\text{\tiny{($+$)}}}+(-1)^{\bar{q}}\frac{V_{y}\eta}{c}\alpha_{\text{\tiny{$\bar{q} L$}}}^{\text{\tiny{($+$)}}}\right|^{2}\label{left-lateral-force}
        \end{align}
    \end{widetext}
    Comparing (\ref{left-lateral-force}) with (\ref{reciprocal-force}), it is clear that the lateral motion affects the four--force (as discussed for the scattering matrix) through mixing the field amplitudes with different polarizations, and through Doppler shifting the frequency and wave--vector within the reflection and transmission coefficients themselves.
    \par
	It is instructive to limit our considerations to situations of practical interest, \textit{i.e.} for relatively small velocities. Upon expanding (\ref{left-lateral-force}) to first order in \(V_{y}/c\), the reflection and transmission coefficients are expanded as \(\mathcal{R}_{qq}^{\prime}\sim\mathcal{R}_{qq}-V_{y}k_{y}\partial\mathcal{R}_{qq}/\partial\omega\).  One then has for the rate of work done, and normal force
\begin{widetext}
   \begin{multline}
   	\left\langle\frac{dE_{\text{\tiny{M}}}}{dt}\right\rangle\sim\frac{A\epsilon_{0}c\zeta_{x}}{2}\sum_{\{q\}}\bigg[\left(1-|\mathcal{R}_{qq}|^{2}-|\mathcal{T}_{qq}|^{2}\right)|\alpha_{qL}^{(+)}|^{2}+2V_{y}k_{y}\text{Re}\left(\mathcal{R}_{qq}^{\star}\frac{\partial\mathcal{R}_{qq}}{\partial\omega}+\mathcal{T}_{qq}^{\star}\frac{\partial\mathcal{T}_{qq}}{\partial\omega}\right)|\alpha_{qL}^{(+)}|^{2}\\
	-(-1)^{\bar{q}}\frac{2V_{y}\eta}{c}\left(|\mathcal{R}_{qq}|^{2}+|\mathcal{T}_{qq}|^{2}\right)\text{Re}\left(\alpha_{qL}^{(+)}\alpha_{\bar{q}L}^{\star(+)}\right)\bigg]\label{first-order-work}
   \end{multline}
    \begin{multline}
   	\left\langle\frac{d\boldsymbol{P}_{\text{\tiny{M$\perp$}}}}{dt}\right\rangle\sim\frac{A\epsilon_{0}\zeta_{x}^{2}}{2}\sum_{\{q\}}\bigg[\left(1+|\mathcal{R}_{qq}|^{2}-|\mathcal{T}_{qq}|^{2}\right)|\alpha_{qL}^{(+)}|^{2}-2V_{y}k_{y}\text{Re}\left(\mathcal{R}_{qq}^{\star}\frac{\partial\mathcal{R}_{qq}}{\partial\omega}-\mathcal{T}_{qq}^{\star}\frac{\partial\mathcal{T}_{qq}}{\partial\omega}\right)|\alpha_{qL}^{(+)}|^{2}\\
	+(-1)^{\bar{q}}\frac{2V_{y}\eta}{c}\left(|\mathcal{R}_{qq}|^{2}-|\mathcal{T}_{qq}|^{2}\right)\text{Re}\left(\alpha_{qL}^{(+)}\alpha_{\bar{q}L}^{\star(+)}\right)\bigg]\label{first-order-normal-force}
   \end{multline}
\end{widetext}
     In this limit we have, \(\eta=|k_{x}|k_{z}/k_{\parallel}^{2}\).  The lateral force, proportional to the loss term in the round bracket on the right hand side of (\ref{left-lateral-force}), may be cast in the simple form
  \begin{equation}
   	\left\langle\frac{d\boldsymbol{P}_{\text{\tiny{M$\parallel$}}}}{dt}\right\rangle=\frac{\boldsymbol{k}_{\parallel}}{\omega} \ \left\langle\frac{dE_{\text{\tiny{M}}}}{dt}\right\rangle,\label{first-order-lateral-force}
   \end{equation}
   Notice that within both (\ref{first-order-work}) and (\ref{first-order-normal-force}) there are two velocity dependent terms. One represents the change in the reflection and transmission coefficients due to the Doppler shifted frequency response, while the other and less familiar one is an interference term arising from the fact that the four force is sensitive to the phase difference between the two complex polarization amplitudes \(\alpha_{\text{\tiny{$1L$}}}^{\text{\tiny{$(+)$}}}\)  and \(\alpha_{\text{\tiny{$2L$}}}^{\text{\tiny{$(+)$}}}\). 
\par
 This polarization dependent contribution is illustrated in figure~\ref{figure-2a} for a \textit{linear} (electric) polarization as a function of the angle of incidence.  The polarization dependent contribution to the force vanishes to first order in \(V_{y}/c\) for circular polarization.  In figure~\ref{figure-2a}b we illustrate an instance where the contribution to the optical force due to polarization mixing may be enhanced.  In this case we reduce the permittivity to a value less than \(1\) and thereby make the medium sensitive to the difference between s and p polarizations for angles where \(\eta=\cot(\theta)\sin(\phi)\) is large.  However, this method is certainly not unique, and it is likely that a greater enhancement would be possible with something like the polarization sensitive mirrors of~\cite{lousse2004}.
     \par
   This is a peculiar effect indeed and does not appear to have been noticed before.  When the phase difference is equal to an integer multiplied by \(\pi\) this contribution is maximized (either positive or negative), while a \(\pi/2\) phase difference causes this velocity dependent contribution to vanish.   Furthermore, under special circumstances the interference term and that due to the Doppler shift may cancel out so as to yield a velocity independent force.  For a lossless medium this could occur \textit{e.g.} when \(\alpha_{\text{\tiny{$1 L$}}}^{(\text{\tiny{$+$}})}=\alpha_{\text{\tiny{$2 L$}}}^{(\text{\tiny{$+$}})}\), and \(ck_{y}\text{Re}[\mathcal{R}_{qq}^{\star}\partial\mathcal{R}_{qq}/\partial\omega]=(-1)^{\bar{q}}\eta|\mathcal{R}_{qq}|^{2}\).
     \par
     The above findings have a relevance to the recent controversy regarding the existence of a drag component to the Casimir force, discussed in the introduction.  Firstly, if the field has an ill defined phase (e.g. a thermal field, or the quantum vacuum), then the first order effect of polarization mixing should average to zero, while the Doppler shift will not, c.f.~\cite{leonhardt2010a}.  Secondly, the contribution of polarization mixing to the force vanishes when the medium has a degenerate response to the two polarizations, whereas the contribution due to the Doppler shift does not.  Finally, it is worth pointing out that the lateral force due to first order polarization mixing, and that due to the Doppler shift have a different dependence upon the direction of incidence, and so would not be expected to cancel in a general situation.  From (\ref{first-order-work}) and (\ref{first-order-lateral-force}) it can be seen that the velocity dependent force in the \(\hat{\boldsymbol{y}}\) direction has a contribution odd in \(k_{y}\) coming from the term proportional to \(\eta\), and even in \(k_{y}\) from the term proportional to the derivatives of the reflection and transmission coefficients.  An isotropic field will therefore---through the effect of the Doppler shift alone---exert a lateral force on a body that is in motion, so long as that body can absorb photons.  This is related to the phenomenon of \emph{universal drag} that was highlighted a few years ago~\cite{mkrtchian2003}, where a dielectric in a thermal field was generally found to experience a frictional force that is linear in the imaginary part of the dielectric susceptibility.\\
%
%
    \subsection{Motion along the \(x\) axis\label{Vx-section}}
    \par
\begin{figure}[hc]
	\includegraphics[width=7cm]{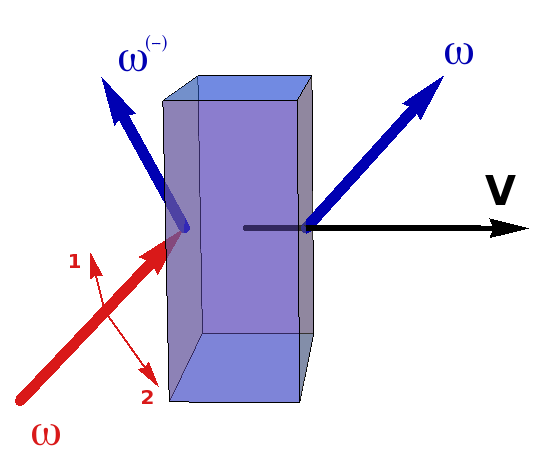}
	\caption{A slab in motion along the \(x\) axis does not mix polarizations, but mixes frequencies.  Incidence from the left, when \(V_{x}>0\) yields a reflected amplitude at a frequency, \(\omega^{(-)}<\omega\), while incidence from the right yields a reflected amplitude at \(\omega^{(+)}>\omega\), which breaks reciprocity.\label{figure-3}}
    \end{figure}

    When a plane-wave is incident onto a surface that moves with velocity \(\boldsymbol{V}=V_{x}\hat{\boldsymbol{x}}\) parallel to the surface normal, we cannot in general use a monochromatic transfer or scattering matrix to determine relationships between field amplitudes.  A plane-wave of given frequency in the lab frame is reflected with a different frequency.  In the slab rest frame the amplitudes of waves moving to the left (\(\alpha_{q\text{\tiny{R}}}^{\text{\tiny{($-$)}}}\)) and moving to the right (\(\alpha_{q\text{\tiny{L}}}^{\text{\tiny{($+$)}}}\)) suffer distinct Doppler shifts.  Therefore for plane waves of fixed  frequency \(\omega\) in the lab frame impinging on both sides of the slab, there are eight rather than four output amplitudes at three different frequencies (\(\omega,\omega^{\text{\tiny{($+$)}}},\omega^{\text{\tiny{($-$)}}}\)).  The result (\ref{four-force-scattering-matrix}) no longer applies and must be modified (see figure \ref{figure-3}).
      \par
      To apply (\ref{four-force-scattering-matrix}) to a slab in motion along \(\hat{\boldsymbol{x}}\), we note that, due to the time averaging of section \ref{tmsec}, that leads to (\ref{averaged-energy-momentum-tensor}), the two different input frequencies observed in rest frame of the slab contribute additively to the energy--momentum tensor, and hence the force~\cite{volume2}.  Therefore, we sum the force due to incidence from the right and left of the slab separately,
      \begin{widetext}
      \begin{align}
          \left\langle\frac{dE^{\prime}_{\text{\tiny{M}}}}{dt^{\prime}}\right\rangle
          &=\frac{A^{\prime}\epsilon_{0}c}{2}\boldsymbol{\alpha}^{\prime\dagger}_{\text{\tiny{(IN)}
          }}\left[\sum_{\{\pm\}}\zeta_{x}^{\prime\text{\tiny{($\pm$)}}}\boldsymbol{\pi}^{\text{\tiny{($\pm$)}}}
          \left(\mathbb{1}_{4}-\boldsymbol{S}^{\prime\dagger}\boldsymbol{S}^{\prime}
          \right)\boldsymbol{\pi}^{\text{\tiny{($\pm$)}}}\right]\boldsymbol{\alpha}^{\prime}_{\text{\tiny{(IN)}}}\nonumber\\
          \left\langle\frac{d\boldsymbol{P}_{\text{\tiny{M}}}^{\prime}}{dt^{\prime}}\right\rangle
          &=\frac{A^{\prime}\epsilon_{0}}{2}\boldsymbol{\alpha}^{\prime\dagger}_{\text{\tiny{(IN)}}
          }\left[\sum_{\{\pm\}}\zeta_{x}^{\prime\text{\tiny{($\pm$)}}}\left(\begin{matrix}
          \boldsymbol{\pi}^{\text{\tiny{($\pm$)}}}
          \left(\boldsymbol{R}-\boldsymbol{S}^{\prime\dagger}\boldsymbol{R}\boldsymbol{S}^{\prime}
          \right)\boldsymbol{\pi}^{\text{\tiny{($\pm$)}}}&\zeta_{x}^{\prime\text{\tiny{($\pm$)}}}\hat{
          \boldsymbol{x}}\\
          \boldsymbol{\pi}^{\text{\tiny{($\pm$)}}}
          \left(\mathbb{1}_{4}-\boldsymbol{S}^{\prime\dagger}\boldsymbol{S}^{\prime}
          \right)\boldsymbol{\pi}^{\text{\tiny{($\pm$)}}}&\zeta_{\parallel}^{\prime\text{\tiny{($\pm$)}}
          }\hat{\boldsymbol{k}}_{\parallel}\end{matrix}\right)\right]\boldsymbol{\alpha}
          ^{\prime}_{\text{\tiny{(IN)}}}\label{four-force-rest-frame}
      \end{align}
      \end{widetext}
      The components of the rest frame scattering matrix, \(\boldsymbol{S}^{\prime}\), are evaluated at the two different rest frame frequencies, \(\omega^{\prime\text{\tiny{($\pm$)}}}=\gamma\left(\omega\mp V_{x}|k_{x}|\right)\) depending on whether they are associated with incidence from the left (+) or from the right (-).  For the reflection coefficients of \(r\)--polarized wave impinging from the left one has \textit{e.g.}  \(\mathcal{R}^{\prime}_{qr}=\mathcal{R}_{qr}(\omega^{\prime\text{\tiny($+$)}})\) and \(\bar{\mathcal{R}}^{\prime}_{qr}=\bar{\mathcal{R}}_{qr}(\omega^{\prime\text{\tiny($-$)}})\) for waves impinging from the right.
      \par
      A Lorentz transformation of (\ref{four-force-rest-frame}) into the laboratory frame gives the dependence of the observed four--force on the velocity of the medium.  The rate of change of energy and normal force on the medium are given by (\(A^{\prime}=A\)),
      \begin{widetext}
      \begin{equation}
          \left\langle\frac{dE_{\text{\tiny{M}}}}{dt}\right\rangle=\frac{A\epsilon_{0}c}{2}\boldsymbol{\alpha}_{\text{\tiny{IN}}}^{\prime\dagger}\sum_{\{\pm\}}\zeta_{x}^{\prime(\pm)}\boldsymbol{\pi}^{(\pm)}\bigg[\left(\mathbb{1}_{4}+\zeta_{x}^{\prime(\pm)}\frac{V_{x}}{c}\boldsymbol{R}\right)-\boldsymbol{S}^{\prime\dagger}\left(\mathbb{1}_{4}+\zeta_{x}^{\prime(\pm)}\frac{V_{x}}{c}\boldsymbol{R}\right)\boldsymbol{S}^{\prime}\bigg]\boldsymbol{\pi}^{(\pm)}\boldsymbol{\alpha}_{\text{\tiny{IN}}}^{\prime}\label{final-moving-force-1}
      \end{equation}\\
      \begin{equation}
      \left\langle\frac{d\boldsymbol{P}_{\perp\text{\tiny{M}}}}{dt}\right\rangle=\frac{A\epsilon_{0}}{2}\boldsymbol{\alpha}_{\text{\tiny{IN}}}^{\prime\dagger}\sum_{\{\pm\}}\zeta_{x}^{\prime(\pm)}\boldsymbol{\pi}^{(\pm)}\bigg[\left(\boldsymbol{R}\zeta_{x}^{\prime(\pm)}+\frac{V_{x}}{c}\mathbb{1}_{4}\right)-\boldsymbol{S}^{\prime\dagger}\left(\boldsymbol{R}\zeta_{x}^{\prime(\pm)}+\frac{V_{x}}{c}\mathbb{1}_{4}\right)\boldsymbol{S}^{\prime}\bigg]\boldsymbol{\pi}^{(\pm)}\boldsymbol{\alpha}_{\text{\tiny{IN}}}^{\prime}\label{final-moving-force-2}
      \end{equation}
      \end{widetext}
      and the lateral force is given by the same expression as in the rest frame (\ref{four-force-rest-frame}).  All that remains is to write all quantities in terms of those in the lab frame.  The directional terms $\zeta$, in particular, transform as,
      \begin{align*}
        \zeta_{x}^{\prime\text{\tiny{($\pm$)}}}&=\frac{c|k_{x}^{\prime\text{\tiny{($\pm$)
        }}}|}{\omega^{\prime\text{\tiny{($\pm$)}}}}=c\left(\frac{|k_{x}|\mp\frac{V_{x}}{c^{
        2}}\omega}{\omega\mp V_{x}|k_{x}|}\right)=\cos(\theta^{\prime\text{\tiny{($\pm$)}}})\\
        \zeta_{\parallel}^{\prime\text{\tiny{($\pm$)}}}&=\frac{ck_{\parallel}^{\prime}}{
        \omega^{\prime\text{\tiny{($\pm$)}}}}=\frac{c
        k_{\parallel}}{\gamma\left(\omega\mp V_{x}|k_{x}|\right)}=\sin(\theta^{\prime\text{\tiny{($\pm$)}}})
      \end{align*}
      In this case the field amplitudes of different polarization do not mix between reference frames, and are related by the diagonal matrix (see appendix \ref{appendix-normal-motion}),
      \[
      		\boldsymbol{\alpha}_{q}^{\prime}=\gamma\left(\mathbb{1}_{2}-\frac{V_{x}|k_{x}|}{\omega}\boldsymbol{\sigma}_{z}\right)\boldsymbol{\alpha}_{q}
      \]
      \par
      As in the previous section we examine the simpler case of incidence from the left onto a medium where \(\boldsymbol{S}_{12}^{\prime}=\boldsymbol{S}_{21}^{\prime}=\boldsymbol{0}\), so that (\ref{final-moving-force-1}--\ref{final-moving-force-2}) become
      \begin{widetext}
      \begin{align}
          \left\langle\frac{dE_{\text{\tiny{M}}}}{dt}\right\rangle&=\frac{
          A\epsilon_{0}c}{2}\left(\frac{c\omega^{\prime}k_{x}^{\prime}}{\omega^{2}}
          \right)\sum_{\{q\}}\left|\alpha_{q\,\text{\tiny{L}}}^{\text{\tiny{($+$)}}}\right|^{2
          }\left[1-\left|\mathcal{R}^{\prime}_{qq}\right|^{2}-\left|\mathcal{T}^{\prime}_{
          qq}\right|^{2}+\frac{V_{x}k_{x}^{\prime}}{\omega^{\prime}}\left(1+\left|\mathcal
          {R}^{\prime}_{qq}\right|^{2}-\left|\mathcal{T}^{\prime}_{qq}\right|^{2}
          \right)\right]\nonumber\\
          \left\langle\frac{d\boldsymbol{P}_{\text{\tiny{M}}}}{dt}\right\rangle&=\frac{
          A\epsilon_{0}}{2}\left(\frac{c\omega^{\prime}k_{x}^{\prime}}{\omega^{2}}
          \right)\sum_{\{q\}}\left|\alpha_{q\,\text{\tiny{L}}}^{\text{\tiny{($+$)}}}\right|^{2
          }\left(\begin{matrix}
          \left[\frac{ck_{x}^{\prime}}{\omega^{\prime}}\left(1+\left|\mathcal{R}^{
          \prime}_{qq}\right|^{2}-\left|\mathcal{T}^{\prime}_{qq}\right|^{2}\right)+\frac{
          V_{x}}{c}\left(1-\left|\mathcal{R}^{\prime}_{qq}\right|^{2}-\left|\mathcal{T}^{
          \prime}_{qq}\right|^{2}\right)\right]&\hat{\boldsymbol{x}}\\
          \frac{c
          k_{\parallel}}{\gamma\omega^{\prime}}\left(1-\left|\mathcal{R}^{\prime}_{qq}\right|^{2
          }-\left|\mathcal{T}^{\prime}_{qq}\right|^{2}\right)&\hat{\boldsymbol{k}}_{\parallel}
          \end{matrix}\right)\label{reciprocal-four-force}
      \end{align}
      \end{widetext}
      where the superscript \((+)\) on the rest frame quantities has been omitted.  The velocity dependence of the force in (\ref{reciprocal-four-force}) has contributions coming from the rest frame loss term, \(1-\left|\mathcal{R}^{\prime}_{qq}\right|^{2}-\left|\mathcal{T}^{\prime}_{qq}\right|^{2}\), the dispersion of the reflection coefficients, and a prefactor of \(c\omega^{\prime}k_{x}^{\prime}/\omega^{2}\) that is \(<1\) for \(V_{x}>0\), and \(>1\) for \(V_{x}<0\).  It is important to note that the force (\ref{reciprocal-four-force}) comprises a lateral component that depends on the loss term \(1-\left|\mathcal{R}^{\prime}_{qq}\right|^{2}-\left|\mathcal{T}^{\prime}_{qq}\right|^{2}\).  For a slab moving along \(\hat{\boldsymbol{x}}\), however, such a loss term is clearly not the only contribution to the rate of loss of energy from the field and---at variance with the case of a laterally moving slab---the lateral force component in (\ref{reciprocal-four-force}) cannot be cast in the simple form  (\ref{first-order-lateral-force}).
      \par
	Retaining only terms linear in \(V_{x}/c\) in (\ref{reciprocal-four-force}) and using the notation of section \ref{Vy-section}, the rate of change of energy and normal force on the slab is,
\begin{widetext}
      \begin{multline}
	\left\langle\frac{dE_{\text{\tiny{M}}}}{dt}\right\rangle\sim\frac{A\epsilon_{0}c}{2}\sum_{\{q\}}\bigg[\zeta_{x}\left(1-|\mathcal{R}_{qq}|^{2}-|\mathcal{T}_{qq}|^{2}\right)+2\zeta_{x}V_{x}k_{x}\text{Re}\left(\mathcal{R}_{qq}\frac{\partial\mathcal{R}_{qq}^{\star}}{\partial\omega}+\mathcal{T}_{qq}\frac{\partial\mathcal{T}_{qq}^{\star}}{\partial\omega}\right)\\
	-\frac{V_{x}}{c}\left(1-(1+2\zeta_{x}^{2})|\mathcal{R}_{qq}|^{2}-|\mathcal{T}_{qq}|^{2}\right)\bigg]|\alpha_{\text{\tiny{$q L$}}}^{(\text{\tiny{$+$}})}|^{2}.\label{first-order-normal-work}
      \end{multline}
      \begin{multline}
      	\left\langle\frac{d\boldsymbol{P}_{\parallel}}{dt}\right\rangle\sim\frac{A\epsilon_{0}\zeta_{\parallel}\hat{\boldsymbol{k}}_{\parallel}}{2}\sum_{\{q\}}\bigg[\left(1-|\mathcal{R}_{qq}|^{2}-|\mathcal{T}_{qq}|^{2}\right)\left(\zeta_{x}-\frac{V_{x}}{c}\right)+2\zeta_{x}V_{x}k_{x}\text{Re}\left(\mathcal{R}_{qq}\frac{\partial\mathcal{R}_{qq}^{\star}}{\partial\omega}+\mathcal{T}_{qq}\frac{\partial\mathcal{T}_{qq}^{\star}}{\partial\omega}\right)\bigg]|\alpha^{(\text{\tiny{$+$}})}_{\text{\tiny{$q L$}}}|^{2}\label{first-order-normal-lateral-force}
      \end{multline}
      \begin{multline}
	\left\langle\frac{d\boldsymbol{P}_{\perp}}{dt}\right\rangle\sim\frac{A\epsilon_{0}\zeta_{x}\hat{\boldsymbol{x}}}{2}\sum_{\{q\}}\bigg[\zeta_{x}\left(1+|\mathcal{R}_{qq}|^{2}-|\mathcal{T}_{qq}|^{2}\right)-2\zeta_{x}V_{x}k_{x}\text{Re}\left(\mathcal{R}_{qq}\frac{\partial\mathcal{R}_{qq}^{\star}}{\partial\omega}-\mathcal{T}_{qq}\frac{\partial\mathcal{T}_{qq}^{\star}}{\partial\omega}\right)\\
	-\frac{V_{x}}{c}\left(1+3|\mathcal{R}_{qq}|^{2}-|\mathcal{T}_{qq}|^{2}\right)\bigg]|\alpha^{(\text{\tiny{$+$}})}_{\text{\tiny{$q L$}}}|^{2}\label{first-order-normal-normal-force}
      \end{multline}
\end{widetext}
      \par
      There are several comments to be made about (\ref{first-order-normal-work}--\ref{first-order-normal-normal-force}).  Firstly, there are no interference terms between TE and TM polarizations due to the motion, as there were in (\ref{first-order-work}--\ref{first-order-normal-force}), and so the phase of the incident field is irrelevant to the optical force.  Secondly, we have the term proportional to \(-(1+3|\mathcal{R}_{qq}|^{2}-|\mathcal{T}_{qq}|^{2})\) in the normal component of the force.  This quantity is always negative, and so the term acts like a friction against the acceleration of the material by the field.  Indeed, for normal incidence onto a dispersionless, lossless medium, this contribution becomes \(-4V_{x}|\mathcal{R}_{qq}|^{2}/c\), which is the radiation friction (or damping) term investigated within~\cite{braginski1967,matsko1996}.  The contribution of the dispersion of the reflection and transmission coefficients within (\ref{first-order-normal-work}--\ref{first-order-normal-normal-force}) is not fundamentally different from that found for lateral motion, and allows for a damping or amplification of the centre of mass motion depending upon sign of the gradients of the reflection and transmission coefficients~\cite{favero2007,horsley2011}.  Note that, unlike the case of lateral motion where the polarization mixing parameter \(\eta\) depends on the plane of incidence, here the role of the plane of incidence of the field is not fundamentally altered from a medium at rest.
%
%
\section{Conclusions}
  \par
  We have developed a general theory of radiation pressure for planar media in terms of a \(4\times4\) scattering matrix, which allows for coupling between the s and p polarizations in transmission through and reflection from a medium.  Applying this theory to the problem of the radiation pressure experienced by a planar medium in motion, it was found that for lateral motion (see figure \ref{figure-2}) the effective scattering matrix of the medium is such that polarizations are coupled (see (\ref{transformed-scattering-matrix})), and something akin to the \(4\times4\) scattering matrix description is unavoidable.  This has unusual consequences for the optical four--force, which then has a velocity dependence that changes according to the phase difference between the amplitudes of the two different polarizations (see (\ref{first-order-work}--\ref{first-order-lateral-force})).  Indeed in the general case (\ref{lateral-result}) the dependence of the force on the rest frame field amplitudes is quite involved.  Meanwhile, for motion parallel to the surface normal there is no mixing of the polarizations, and to the lowest order in \(V/c\) our formalism recovers the known results for moving surfaces~\cite{braginski1967,matsko1996,horsley2011}.  It is indeed possible to apply this formalism to the general case of a medium in motion in an arbitrary direction, for we can consider this as a combination of motion normal and parallel to the plane of incidence.
  \par
  One clear advantage of this formalism is its relative simplicity.  Work on the problem of quantum friction often encounters very cumbersome formulae on which there seems to be no general agreement (e.g.~\cite{volokitin2010}).  The fact that lateral motion can be thought of as having two effects---a unitary transformation of the transfer/scattering matrix, and a Doppler shift of the frequency within the reflection \& transmission coefficients is therefore advantageous, and it should be possible to apply this formalism to the problem in a straightforward way through a generalization of~\cite{genet2003}.

\acknowledgments
    One of us (SARH) would like to thank the EPSRC for financial support.  This work was also supported by the CRUI-British Council Programs ``Atoms and Nanostructures" and ``Metamaterials", and the IT09L244H5 Azione Integrata MIUR grant.
%
%
\appendix
%
%
    \section{The Lorentz transformation of polarization\label{appendix-a}}
    \par
    Here we examine how polarization transforms between reference frames.  We consider motion of two kinds; either parallel to the surface normal (motion along \(\hat{\boldsymbol{x}}\)), or orthogonal to the surface normal (motion along \(\hat{\boldsymbol{y}}\)).
    \subsection{Motion along \(\hat{\boldsymbol{x}}\)\label{appendix-normal-motion}}
    \par
    When a planar medium moves in a direction parallel to the surface normal, then after scattering from the medium, the fields (\ref{electric-magnetic-fields}) will no longer be monochromatic.  However, the phase factors within (\ref{electric-magnetic-fields}) will take the same value in any inertial reference frame.  Therefore after applying the transformation
    \begin{align}
    	E_{x}^{\prime}&=E_{x}\nonumber\\
	E_{y}^{\prime}&=\gamma(E_{y}-V_{x}B_{z})\nonumber\\
	E_{z}^{\prime}&=\gamma(E_{z}+V_{x}B_{y})\label{x-trans}
    \end{align}
    we obtain the equations
    \begin{equation}
    	\hat{e}_{2x}^{\prime(\pm)}\alpha_{2}^{\prime(\pm)}=\hat{e}_{2x}^{(\pm)}\alpha_{2}^{(\pm)}\label{x-trans-1}
    \end{equation}
    and
    \begin{multline}
    	\hat{e}_{1y}^{\prime(\pm)}\alpha_{1}^{\prime(\pm)}+\hat{e}_{2y}^{\prime(\pm)}\alpha_{2}^{\prime(\pm)}=\\[7pt]
	\gamma\bigg[\left(\hat{e}_{1y}^{(\pm)}-\frac{V_{x}}{c}\hat{e}_{2z}^{(\pm)}\right)\alpha_{1}^{(\pm)}
	+\left(\hat{e}_{2y}^{(\pm)}+\frac{V_{x}}{c}\hat{e}_{1z}^{(\pm)}\right)\alpha_{2}^{(\pm)}\bigg]\label{x-trans-2}
    \end{multline}
    with the third of (\ref{x-trans}) holding identically, given (\ref{x-trans-1}--\ref{x-trans-2}).  We then insert the components of the polarization unit vectors which from (\ref{pol1}--\ref{pol2}) are; \(\hat{e}_{2x}^{(\pm)}=ck_{\parallel}/\omega\); \(\hat{e_{1z}}^{(\pm)}=k_{y}/k_{\parallel}\); \(\hat{e}_{2z}^{(\pm)}=\mp c|k_{x}|k_{z}/\omega k_{\parallel}\); \(\hat{e}_{1y}^{(\pm)}=-k_{z}/k_{\parallel}\); and \(\hat{e}^{(\pm)}_{2y}=\mp c|k_{x}|k_{y}/\omega k_{\parallel}\).  For (\ref{x-trans-1}) this gives
    \begin{equation}
    	\alpha^{\prime(\pm)}_{2}=\left(\frac{\omega^{(\pm)}}{\omega}\right)\alpha_{2}^{(\pm)}\label{normal-result-1}
    \end{equation}
    where \(\omega^{(\pm)}=\gamma(\omega\mp V|k_{x}|)\).  Note that we do not consider the case where \(k_{x}\) changes sign between reference frames, although this is interesting.  Inserting (\ref{normal-result-1}) into (\ref{x-trans-2}) along with the components of the unit polarization vectors, we find
    \begin{equation}
    	\alpha_{1}^{\prime(\pm)}=\left(\frac{\omega^{(\pm)}}{\omega}\right)\alpha_{1}^{(\pm)}\label{normal-result-2}
    \end{equation}
    The field amplitudes consequently transform as
    \begin{equation}
    	\boldsymbol{\alpha}_{q}^{\prime}=\gamma\left(\mathbb{1}_{2}-\frac{V_{x}|k_{x}|}{\omega}\boldsymbol{\sigma}_{z}\right)\boldsymbol{\alpha}_{q}\label{normal-result}
    \end{equation}
    between transversely moving inertial frames.  In this case the polarization of the wave does not change between frames, the amplitude being either increased or decreased according to the ratio of the rest frame and the laboratory frame frequency.
%
%
     \subsection{Motion along \(\hat{\boldsymbol{y}}\)\label{appendix-lateral-motion}}
    \par
    In the case of a medium moving laterally as in figure \ref{figure-2}, the frequency \(\omega\) is conserved.  Applying the transformation
    \begin{align}
    	E_{x}^{\prime}&=\gamma\left(E_{x}+V_{y}B_{z}\right)\nonumber\\
	E_{y}^{\prime}&=E_{y}\nonumber\\
	E_{z}^{\prime}&=\gamma\left(E_{z}-V_{y}B_{x}\right)\label{lateral-trans}
    \end{align}
    to (\ref{electric-magnetic-fields}) along with the equivalent primed quantities leads to the equations
    \begin{equation}
    	\hat{e}_{2x}^{\prime(\pm)}\alpha_{2}^{\prime(\pm)}=\gamma\left[\alpha_{2}^{(\pm)}\left(\hat{e}_{2x}^{(\pm)}-\frac{V_{y}}{c}\hat{e}_{1z}^{(\pm)}\right)+\frac{V_{y}}{c}\hat{e}_{2z}^{(\pm)}\alpha_{1}^{(\pm)}\right]\label{pol-trans-1}
    \end{equation}
    and
    \begin{equation}
    	\hat{e}_{1y}^{\prime(\pm)}\alpha_{1}^{\prime(\pm)}+\hat{e}_{2y}^{\prime(\pm)}\alpha_{2}^{\prime(\pm)}=\hat{e}_{1y}^{(\pm)}\alpha_{1}^{(\pm)}+\hat{e}_{2y}^{(\pm)}\alpha_{2}^{(\pm)}\label{pol-trans-2}
    \end{equation}
    The third equation in (\ref{lateral-trans}) holds identically when we apply (\ref{pol-trans-1}--\ref{pol-trans-2}).  We insert the components of the unit polarization vectors given in appendix \ref{appendix-normal-motion}.  In the case of (\ref{pol-trans-1}) this gives
    \[
    	\alpha_{2}^{\prime(\pm)}=\frac{\gamma\omega^{\prime}}{\omega k_{\parallel}k_{\parallel}^{\prime}}\left[\alpha^{(\pm)}_{2}\left(k_{\parallel}^{2}-\frac{V_{y}k_{y}\omega}{c^{2}}\right)\mp\frac{V_{y}}{c}|k_{x}|k_{z}\alpha_{1}^{(\pm)}\right]
    \]
    Defining \(\eta=|k_{x}|k_{z}/(k_{\parallel}^{2}-V_{y}k_{y}\omega/c^{2})\) and noting that
    \[
    	\sqrt{1+V_{y}^2\eta^{2}/c^{2}}=\frac{k_{\parallel}k_{\parallel}^{\prime}}{\gamma(k_{\parallel}^{2}-V_{y}k_{y}\omega/c^{2})}
    \]
    we find the transformation formula for the p--polarized amplitudes
    \begin{equation}
    	\alpha_{2}^{\prime(\pm)}=\left(\frac{\omega^{\prime}}{\omega}\right)\left(\frac{\alpha^{(\pm)}_{2}\mp\frac{V_{y}\eta}{c}\alpha_{1}^{(\pm)}}{\sqrt{1+V_{y}^2\eta^{2}/c^{2}}}\right)\label{p-trans}
    \end{equation}
    Inserting (\ref{p-trans}) and the components of the unit vectors into (\ref{pol-trans-2}) then gives the transformation formula for s--polarized amplitudes
    \begin{equation}
    	\alpha_{1}^{\prime(\pm)}=\left(\frac{\omega^{\prime}}{\omega}\right)\left(\frac{\alpha^{(\pm)}_{1}\pm\frac{V_{y}\eta}{c}\alpha_{2}^{(\pm)}}{\sqrt{1+V_{y}^2\eta^{2}/c^{2}}}\right)\label{s-trans}
    \end{equation}
    In matrix notation, (\ref{p-trans}) and (\ref{s-trans}) become
    \begin{equation}
    	\boldsymbol{\alpha}^{\prime}=\left(\frac{\omega^{\prime}}{\omega}\right)\boldsymbol{M}\boldsymbol{\cdot}\boldsymbol{\alpha}
    \end{equation}
    where
    \begin{equation}
    	\boldsymbol{M}=\frac{1}{\sqrt{1+V_{y}^2\eta^{2}/c^{2}}}\left(\begin{matrix}\mathbb{1}_{2}&\frac{V_{y}\eta}{c}\boldsymbol{\sigma}_{z}\\
								-\frac{V_{y}\eta}{c}\boldsymbol{\sigma}_{z}&\mathbb{1}_{2}\end{matrix}\right)
    \end{equation}
    This unitary matrix, \(\boldsymbol{M}\) describes the transformation of the polarization of a plane wave between laterally moving reference frames.
    \bibliography{rppc-refs}
\end{document}